\documentclass[11pt,a4paper,preprintnumbers,nofootinbib,superscriptaddress]{article}
\usepackage{jheppub}
\newcommand{\calG}{\mathcal{G}}
\newcommand{\tcalG}{\widetilde{\mathcal{G}}}

\begin{document}

\title{The most general second-order field equations of bi-scalar-tensor
theory in four dimensions}

\author[a]{Seiju~Ohashi,}
\affiliation[a]{Cosmophys Group, IPNS KEK, 1-1 Oho, Tsukuba 305-0801, Japan}
\emailAdd{sohashi"at"post.kek.jp}

\author[b]{Norihiro~Tanahashi,}
\affiliation[b]{DAMTP, Centre for Mathematical Sciences, University of Cambridge, Wilberforce Road, Cambridge CB3 0WA, UK}
\emailAdd{N.Tanahashi"at"damtp.cam.ac.uk}

\author[c]{Tsutomu~Kobayashi}
\affiliation[c]{Department of Physics, Rikkyo University, Toshima, Tokyo 175-8501, Japan}
\emailAdd{tsutomu"at"rikkyo.ac.jp}

\author[d]{and Masahide~Yamaguchi}
\affiliation[d]{Department of Physics, Tokyo Institute of Technology, Tokyo 152-8551, Japan}
\emailAdd{gucci"at"phys.titech.ac.jp}

\abstract{
The Horndeski theory is known as the most general scalar-tensor theory
with second-order field equations. In this paper, we explore the
bi-scalar extension of the Horndeski theory.  Following Horndeski's
approach, we determine all the possible terms appearing in the
second-order field equations of the bi-scalar-tensor theory.  We compare
the field equations with those of the generalized multi-Galileons, and
confirm that our theory contains new terms that are not included in the
latter theory. We also discuss the construction of the Lagrangian
leading to our most general field equations.
}

\preprint{KEK-TH-1826, RUP-15-11}
\maketitle

\section{Introduction}

The inflationary scenario \cite{inflation} is now regarded as a necessary
ingredient of modern cosmology because the recent observations of cosmic
microwave background anisotropies
\cite{Bennett:2012zja,Hinshaw:2012aka,Ade:2013ktc,Ade:2013uln} strongly
suggest the epoch of inflationary expansion in the early Universe. Since
inflation is typically driven by scalar field(s), scalar-tensor theories
provide a firm framework to study the dynamics of inflation. On the
other hand, the unknown energy called dark energy is shown to be
dominant in the present Universe
\cite{Schmidt:1998ys,Riess:1998cb,Perlmutter:1998np} and might be
understood as an outcome of infra-red modification of
gravity. Scalar-tensor theories provide powerful tools to realize such a
modification, and their phenomenological features have been studied
intensively to confront them against observations.  Although a plethora
of inflationary models and modified gravity theories have been proposed
thus far, unfortunately, we have not yet succeeded in finding the real
theory and have kept seeking for it. We have two options to address this
problem. The first one is to pursue the ultimate (real) theory on the
basis of theoretical consistencies, which is often called top-down
approach. The other is to construct a framework of theories as general
as possible, which is called bottom-up approach. Of course, though both
of approaches are complementary, one of the merits to take the latter
approach is to give a unified understanding of various models proposed
individually. Another is that one can easily pin down or constrain
models once characteristic observational results would be reported.

The Horndeski theory \cite{Horndeski:1974wa} provides a typical working
example of the latter approach because it is the most general
single scalar-tensor theory with second-order field equations. It is
shown \cite{Kobayashi:2011nu} that the Horndeski theory is equivalent to
the generalized Galileon \cite{Deffayet:2011gz}. In fact, almost all of
the inflationary models with single inflaton proposed so far can be
described by this theory in a unified manner. Various aspects of single
field inflationary models have been studied in this
framework~\cite{Kobayashi:2011nu}, which is useful for us to constrain
the models from observational results.

In this paper, we take the latter approach and try to extend
the Horndeski theory, which includes
only one scalar degree of freedom, to the bi-scalar case.\footnote{Another direction is to consider a
scalar-vector theory instead of a scalar-tensor theory. Motivated by the
earlier work of Horndeski~\cite{Horndeski:1976gi}, construction of
the most general vector theory with second order field equations on flat
space, called the vector-Galileon theory, was attempted in
Ref.~\cite{Deffayet:2013tca}. Recently, the scalar-tensor theory with
higher order equations of motion without introducing the ghost was
proposed as well \cite{Gleyzes:2014dya,Gao:2014soa}. See also~\cite{Zumalacarregui:2013pma}
for yet another possible way of extending the Horndeski theory.}
Although no
observational results suggest the presence of multiple (light) scalars
during inflation, our framework can clarify the essential difference
between single and multiple scalar cases, and is useful for further
constraining or detecting the presence of multiple scalar degrees of
freedom from observations. Such an attempt to extend the Horndeski
theory has already been discussed. First of all, multi-field Galileon
theory was proposed in the flat spacetime
\cite{Padilla:2010de,Padilla:2010tj,Hinterbichler:2010xn,Padilla:2010ir}. Later,
the covariantization of this multi-field Galileon theory, called
generalized multi-Galileon, was considered~\cite{Padilla:2012dx} and
conjectured that the theory would correspond to the multi-field
extension of the Horndeski theory, that is, the most general multi-field
scalar-tensor theory with second order equations of motion. Though the
multi-field Galileon theory in the flat spacetime is proven to be the
most general multiple-scalar field theory in the flat space-time with
second order scalar equations of motion \cite{Sivanesan:2013tba}, it was
shown that the generalized multi-Galileon is {\it not} the most general
theory \cite{Kobayashi:2013ina} because this theory does not contain the
multi-DBI inflation
models~\cite{RenauxPetel:2011dv,RenauxPetel:2011uk,Koyama:2013wma,Langlois:2008wt,Langlois:2008qf},
in particular the double-dual Riemann term appearing in these
models. Motivated by these considerations, in this paper, we try to
construct a multi-field extension of the Horndeski theory. Especially we
focus on the extension to the bi-scalar case as a first step.

The organization of this paper is as follows.  In
Sec.~\ref{Sec:EoM}, we construct the field equations of the most
general two-scalar tensor theory with second order field equations
following Horndeski's procedure. This section is the main part of this
paper.  In Sec.~\ref{Sec:multiG}, we compare our theory with the
generalized multi-Galileon theory.  We show that terms which are missing
in generalized multi-Galileon theory are actually contained in our
theory. In Sec.~\ref{Sec:L}, we comment on the construction of the
Lagrangian corresponding to the field equations we obtain.  Finally we
summarize our paper and discuss the results in
Sec.~\ref{Sec:Discussion}.

\subsection*{Notations and conventions}

Before closing the introduction, we summarize the notations and
conventions used throughout this paper. We consider a four-dimensional
spacetime with a metric $g_{ab}$ and two scalar fields $\phi^I$ with
$I=1,2$.  Following Ref.~\cite{Horndeski:1974wa}, derivatives of
$g_{ab}$ and $\phi^I$ with respect to the coordinates $x^a$ are denoted
as
\begin{align}
g_{ab,c}\equiv \frac{\partial g_{ab}}{\partial x^c}, \quad
\phi^I_{,a}\equiv \frac{\partial \phi^I}{\partial x^a},
\end{align}
respectively.
We denote the covariant derivative of $\phi^I$ with respect to $g_{ab}$ and its scalar product respectively as
\begin{align}
 \phi^I_{|a}\equiv \nabla_a \phi^I ,
\qquad
X^{IJ}\equiv -\frac{1}{2}\phi^{I}_{|a}\phi^{J|a},
\end{align}
where $X^{IJ}$ is symmetric in $I$ and $J$.
We use a strike `` $|$ '' also as a separator
in (anti-)symmetrization. For example, $[I|JK,L|M]$ stands for anti-symmetrization of $I$ and $M$.
Partial derivatives of a function
$
A^{a\dots b}(g, \partial g, \partial^2 g,\phi^I ,\partial \phi^I, \partial^2 \phi^I )$
are expressed as
\begin{align}
A^{a\dots b;cd}& \equiv \frac{\partial A^{a\dots b}}{\partial g_{cd}}, &
A^{a\dots b;cd,e}& \equiv \frac{\partial A^{a\dots b}}{\partial g_{cd,e}}, &
A^{a\dots b;cd,ef}& \equiv \frac{\partial A^{a\dots b}}{\partial g_{cd,ef}},\notag \\
A^{a\dots b;}_{\ \ \ \ \ I}& \equiv \frac{\partial A^{a\dots b}}{\partial \phi^I}, &
A^{a\dots b;c}_{\ \ \ \ \ I}& \equiv \frac{\partial A^{a\dots b}}{\partial \phi^I_{,c}},&
A^{a\dots b;cd}_{\ \ \ \ \ I}& \equiv \frac{\partial A^{a\dots b}}{\partial \phi^I_{,cd}},
\end{align}
and partial derivatives of a function $A(\phi^I,X^{JK})$ are expressed as
\begin{align}
A_{,I} \equiv \frac{\partial A}{\partial \phi^I}
,
\qquad
A_{,IJ} \equiv 
\frac12\left(
\frac{\partial A}{\partial X^{IJ}}
+ \frac{\partial A}{\partial X^{JI}}
\right)
.
\end{align}
In the equations of motion and the Lagrangian, we use the generalized Kronecker delta defined by
\begin{align}
\delta^{i_1 \ldots i_n}_{j_1 \ldots j_n} \equiv 
n! \, \delta^{i_1}_{[j_1}\ldots \delta^{i_n}_{j_n]}
,
\qquad
\delta^{IK}_{JL} \equiv 2 \delta^I_{[J}\delta_{L]}^K .
\end{align}
Repeated indices are summed over $a=0,1,2,3$ and $I=1,2$.

\section{Construction of the most general equations of motion}
\label{Sec:EoM}

The first step of the construction of the most general scalar-tensor
theory of Ref.~\cite{Horndeski:1974wa} is to work out the most general
equations of motion that are of second order in derivatives and
compatible with the general covariance. In this section, we generalize
this construction to the case with two scalar fields.

\subsection{Assumptions}

The assumptions imposed on the theory we are going to construct are summarized as follows.
\begin{enumerate}
\item The theory has a Lagrangian scalar density, $\mathcal{L}$.
\item The Lagrangian scalar density, $\mathcal{L}$, is composed of a metric,
two scalar fields, and their derivatives up to arbitrary order:
\begin{equation}
\mathcal{L}=\mathcal{L}\left( g_{ab},g_{ab,c},g_{ab,cd},\dots ;\phi^I , \phi^I_{,a},\phi^I_{,ab},\dots \right),
\end{equation}
where $I=1,2.$
\item Field equations are composed of the metric, the two scalar fields, and their derivatives up to second order:
\begin{align}
0 = \frac{ \delta \mathcal{L}}{\delta g_{ab}}
&=\sqrt{-g} \, \calG{}^{ab}\left( g_{cd},g_{cd,e},g_{cd,ef};\phi^J , \phi^J_{,c},\phi^J_{,cd}\right),\\
0 =\frac{ \delta \mathcal{L}}{\delta \phi^I} &=\sqrt{-g} \,
\mathcal{E}_I \left( g_{ab},g_{ab,c},g_{ab,cd};\phi^J , \phi^J_{,a},\phi^J_{,ab}\right),
\end{align}
where $\delta \mathcal{L}/\delta A $ is the variation of $\mathcal{L}$ with respect to a field $A$. 
\end{enumerate}

Let us consider the variation of the
action under an infinitesimal coordinate transformation, $x^a \to x^a+\xi^a$, which is given by
\begin{align}
\delta \int d^4x \, \mathcal{L}
=2\int d^4x \sqrt{-g} \biggl( \nabla_b \calG{}^{ab}-\frac{1}{2}
\mathcal{E}_I\nabla^a \phi^I \biggr) \xi_a. \label{ChangeL}
\end{align}
It follows from
the assumption of $\mathcal{L}$ being a scalar density that
Eq.~(\ref{ChangeL}) vanishes identically, implying
that the integrand itself vanishes since $\xi_a$ may be chosen arbitrarily. 
Thus, the identity 
\begin{equation}
\nabla_b \calG{}^{ab}=\frac{1}{2}
\mathcal{E}_I\nabla^a \phi^I \label{G-Bianchi}
\end{equation}
holds. In the case of pure Einstein gravity
this identity reduces to the well-known contracted Bianchi identity, $\nabla_aG^{ab}=0$.
In this sense, the identity~(\ref{G-Bianchi}) may be regarded as a generalization of the Bianchi identity.
We are trying to construct the most general bi-scalar-tensor theory with second-order field equations
using the identity~(\ref{G-Bianchi}). 
Because both $\calG{}^{ab}$ and $\mathcal{E}_I$ are assumed to be of second order in derivatives,
the left-hand side of Eq.~(\ref{G-Bianchi}) would yield third derivatives
in general while the right-hand side contains at most second derivatives.
This indicates that the left-hand side of Eq.~(\ref{G-Bianchi}) must be
free from third derivatives.

The construction of the most general second-order equations of motion
for bi-scalar-tensor theories is divided into two parts:
we first determine the most general second-order rank-2 tensor whose divergence remains of
second order. After that, we impose the identity~(\ref{G-Bianchi}) on the rank-2 tensor to constrain its form.

\subsection{Second-order rank-2 tensor whose divergence is of second order}
\label{Sec:xi}

In this subsection we construct the most general second-order rank-2 tensor, $\tcalG{}^{ab}$,
whose divergence is also of second order. The conditions that $\nabla_a\tcalG{}^{ab}$ has no third derivatives
can be expressed as
\begin{align}
&\frac{\partial \, \nabla_b\tcalG{}^{ab}}{\partial g_{cd,efg}}=0 \label{No3rdG},\\
&\frac{\partial \, \nabla_b\tcalG{}^{ab}}{\partial \phi^I_{,cde}}=0.\label{No3rdphi}
\end{align}
Using the chain rule, $\nabla_b\tcalG{}^{ab}$ is rewritten as
\begin{align}
\nabla_b\tcalG{}^{ab}&= 
\tcalG{}^{ab;cd,ef}g_{cd,efb}+ \tcalG{}^{ab;cd,e}g_{cd,eb}+\tcalG{}^{ab;cd}g_{cd,b}
+\tcalG{}^{ab;cd}_{\ \ \ I}\phi^I_{,cdb}+ \tcalG{}^{ab;c}_{\ \ \ I}\phi^I_{,cb}+ \tcalG{}^{ab;}_{\ \ \ I}\phi^I_{,b}
+\tcalG{}^{bc}\Gamma^a_{bc}+\tcalG{}^{ab}\Gamma^c_{bc},
\label{chain}
\end{align}
where $\Gamma_{bc}^a$ are the Christoffel symbols.
With the help of Eq.~(\ref{chain}), we can show that the conditions (\ref{No3rdG}) and (\ref{No3rdphi}) are equivalent to 
\begin{align}
&\tcalG{}^{ab;cd,ef}+\tcalG{}^{ae;cd,fb}+\tcalG{}^{af;cd,be}=0,\label{No3rdG1}\\
&\tcalG{}^{ab;cd}_{\ \ \ I}+\tcalG{}^{ac;db}_{\ \ \ I}+\tcalG{}^{ad;bc}_{\ \ \ I}=0\label{No3rdphi1},
\end{align}
respectively. From the ``invariance identity'' (see Refs.~\cite{Rund,Plessis}), we have
\begin{equation}
\tcalG{}^{ab;cd,ef}+\tcalG{}^{ab;ce,fd}+\tcalG{}^{ab;cf,de}=0.\label{TensorCondi}
\end{equation}
By repeated use of Eqs.~(\ref{No3rdG1}),~(\ref{No3rdphi1}), and~(\ref{TensorCondi}), we obtain
\begin{align}
&\tcalG{}^{ab;cd,ef}=\tcalG{}^{cd;ab,ef}=\tcalG{}^{ef;cd,ab},\\
&\tcalG{}^{ab;cd}_{\ \ \ I}=\tcalG{}^{cd;ab}_{\ \ \ I}.\label{No3rdG2}
\end{align}

For convenience, we now introduce the notion of {\em property S}
following Ref.~\cite{Horndeski:1974wa}. 
A quantity $A^{a_1a_2\dots a_{2n-1}a_{2n}}$ is said to have property~$S$ if it satisfies the following conditions: 
(i) it is symmetric in $(a_{2\ell -1},a_{2\ell})$ for $\ell=1,2,\dots, n$;
(ii) it is symmetric under the
interchange of any two pairs $(a_{2\ell -1},a_{2\ell})$ and $(a_{2m-1},a_{2m})$ for $\ell ,m=1,2,\dots ,n$;
(iii) it vanishes if any three of four indices,
$(a_{2\ell -1},a_{2\ell})$ and $(a_{2m-1},a_{2m})$ for $\ell ,m=1,2,\dots ,n$, are symmetrized.
It is shown in Corollary 2.1 of Ref.~\cite{Horndeski:1974wa}
that $A^{a_1a_2\dots a_{2n-1}a_{2n}}$ vanishes if $A^{a_1a_2\dots a_{2n-1}a_{2n}}$
has property $S$ and $n > 4$ in four-dimensional spacetime.

Let us introduce the quantity $B^{a_1a_2\dots a_{4n+2m+1}a_{4n+2m+2}}$ defined by
\begin{align}
B^{a_1a_2\dots a_{4n+2m+1}a_{4n+2m+2}}\equiv \Pi_{i=1}^n\Bigg(\frac{\partial }{\partial g_{a_{4i-1}a_{4i},a_{4i+1}a_{4i+2}}} \Bigg) \Pi_{j=1}^m \Bigg( \frac{\partial }{\partial \phi^{I_j}_{,a_{4n+2j+1}a_{4n+2j+2}}}\Bigg) \tcalG{}^{a_1a_2} .
\end{align}
This is a $n$-th derivative with respect to $g_{a_ia_{i+1},a_{i+2}a_{i+3}}$
and $m$-th derivative with respect to $\phi^I_{,a_ia_{i+1}}$ of $\tcalG{}^{ab}$.
Using Eqs.~(\ref{No3rdG1})--(\ref{No3rdG2}), one can easily check
that $B^{a_1a_2\dots a_{4n+2m+1}a_{4n+2m+2}}$ has property $S$.
Then, Corollary $2.1$ of Ref.~\cite{Horndeski:1974wa} implies that
$B^{a_1a_2\dots a_{4n+2m+1}a_{4n+2m+2}}$ vanishes for $2n+m \geq 4$, leading to
the following three sets of identities:
\begin{align}
&\frac{\partial }{\partial g_{cd,ef}}\frac{\partial }{\partial g_{ij,kl}}\tcalG{}^{ab}=0,\label{DiffIdentity1} \\
&\frac{\partial }{\partial g_{cd,ef}}\frac{\partial }{\partial \phi^I_{,ij}}\frac{\partial }{\partial \phi^J_{,kl}}\tcalG{}^{ab}=0,\label{DiffIdentity2}\\
&\frac{\partial }{\partial \phi^I_{,cd}}\frac{\partial }{\partial \phi^J_{,ef}}
\frac{\partial }{\partial \phi^K_{,ij}}\frac{\partial }{\partial \phi^L_{,kl}}\tcalG{}^{ab}=0.\label{DiffIdentity3}
\end{align}

By integrating Eqs.~(\ref{DiffIdentity1})--(\ref{DiffIdentity3}), we can
determine the form of the gravitational field equations.
First, integrating Eq.~(\ref{DiffIdentity1}) yields
\begin{equation}
\tcalG{}^{ab}=\widetilde{\xi}^{abcdef}g_{cd,ef}+\widetilde{\xi}^{ab} 
=\xi^{abcdef}R_{cdef}+\xi^{ab},
\label{IntegGG}
\end{equation}
where $\xi^{abcdef}$ and $\xi^{ab}$ are functions of
$g_{ab}, g_{ab,c}, \phi^I, \phi^I_{,a}$, and $\phi^I_{,ab}$.
Note that we have used the identity
\begin{align}
\tilde{\xi}^{abcdef}g_{cd,ef}=\frac{2}{3}\tilde{\xi}^{abcdef}R_{ecdf}+\bar{\xi}^{ab}
\end{align}
at the second equality of Eq.~(\ref{IntegGG}), where $\bar{\xi}^{ab}$ are functions
of $g_{ab}, g_{ab,c}, \phi^I, \phi^I_{,a}$ and
$\phi^I_{,ab}$.
It can be seen that $\xi^{abcdef}$ and $\xi^{ab}$ have property~$S$.
Substituting Eq.~(\ref{IntegGG}) into Eq.~(\ref{DiffIdentity2}) and integrating it,
we obtain
\begin{align}
\tcalG{}^{ab}&=
\xi^{abcdefgh}_IR_{cdef}\phi^I_{|gh}+\xi^{abcdef}R_{cdef}
+\xi^{ab}, \label{preeom1}
\end{align}
where $\xi^{abcdefgh}$ and $\xi^{abcdefgh}_{I}$ are
functions of $g_{ab}, g_{ab,c}, \phi^I$,
and $\phi^I_{,a}$, while
$\xi^{ab}$ are functions of $g_{ab}$, $g_{ab,c}$, $\phi^I$, $\phi^I_{,a}$ and $\phi^I_{,ab}$. 
Here again it can be seen that $\xi^{abcdefgh}$, $\xi^{abcdefgh}_{I}$ and $\xi^{ab}$ have property~$S$.
Repeating the same procedure and integrating Eq.~(\ref{DiffIdentity3}) give
\begin{align}
\tcalG{}^{ab}&=
\xi^{abcdefgh}_IR_{cdef}\phi^I_{|gh}
+
\xi^{abcdefgh}_{IJK}\phi^I_{|cd}\phi^J_{|ef}\phi^K_{|gh} 
+\xi^{abcdef}R_{cdef} 
+
\xi^{abcdef}_{IJ}\phi^I_{|cd}\phi^J_{|ef}
+
\xi^{abcd}_I\phi^I_{|cd}+\xi^{ab}, 
\label{preeom2}
\end{align}
where all of the above $\xi$ tensors
are composed of $g_{ab}, g_{ab,c}, \phi^I$, and $\phi^I_{,a}$, and have property $S$.
Although our final goal is to determine the most general equations of motion
for the bi-scalar-tensor theory, the equations given up to this point hold irrespective
of the number of the scalar fields.

Our remaining task in this subsection is to construct explicitly all the
possible $\xi$ tensors that have property~$S$ and are composed of
$g_{ab}, g_{ab,c}, \phi^I$, and $\phi^I_{,a}$.  For this purpose, we can
use $\phi^{I|a}$, $g^{ab}$, and the totally antisymmetric tensor
$\varepsilon^{abcd}$ as building blocks, from which the $\xi$ tensors
are built by taking their products and linear combinations
appropriately.  There is no elegant way, and what we will do is to
exhaust all the possible combinations of those building blocks yielding
the $\xi$ tensors.  Let us begin with the simplest one, $\xi^{ab}$.  It
is not difficult to find that the following one is the most general
symmetric rank-2 tensor composed of $\phi^I, \phi^I_{|a}, g^{ab}$, and
$g^{ab,c}$:
\begin{align}
\xi^{ab}=a(\phi^I,X^{JK})g^{ab}+
b_{IJ}(\phi^I,X^{JK})\phi^{I|a}\phi^{J|b},
\label{xiab}
\end{align} 
where $a(\phi^I,X^{JK})$ and $b^{IJ}(\phi^I,X^{JK})$ are arbitrary functions of $\phi^I$ and $X^{JK}$,
and $b^{IJ}$ has the symmetric property $b^{IJ}=b^{JI}$.\footnote{One
might consider other rank-2 tensors such as $\varepsilon^{abcd}\phi^I_{|c}\phi^J_{|d}$,
but this tensor is excluded because it is not symmetric in $a, b$.}
Here, we have used, for the first time in this derivation, the assumption that the
number of the scalar fields is two, which greatly simplifies the
expression of $\xi$ tensors and the following procedure. Without this restriction we would have for example the
term such as $c_{IJKL}\left( \varepsilon^{acde}\phi^{I}_{|c}\phi^{J}_{|d}\phi^{K}_{|e} \phi^{L}{}^{|b} +\varepsilon^{bcde}\phi^{I}_{|c}\phi^{J}_{|d}\phi^{K}_{|e} \phi^{L}{}^{|a} \right)$ in $\xi^{ab}$, where $c_{IJKL}$ is arbitrary functions of $\phi^I$ and $X^{IJ}$, and totally anti-symmetric in $I,J$ and $K$.
In a similar manner, we work out $\xi^{abcd}_I$:
\begin{align}
\xi^{abcd}_I&=a_I\left( g^{ac}g^{bd}+g^{ad}g^{bc}-2g^{ab}g^{cd} \right) \notag \\
                   &~~ +
b_{IJK}\left[ g^{ac}\phi^{J|b}\phi^{K|d}+ g^{ad}\phi^{J|b}\phi^{K|c}
+ g^{bd}\phi^{J |a}\phi^{K |c}+ g^{bc}\phi^{J|a}\phi^{K|d}-2 \left(
g^{ab}\phi^{J|c}\phi^{K|d}+g^{cd}\phi^{J|a}\phi^{K|b}\right)\right] \notag \\
                  &~~ +
c_{IJKLM}\left( 
  \phi^{J |a}\phi^{K|c}\phi^{L|b}\phi^{M|d}
+ \phi^{J|a}\phi^{K|d}\phi^{L|b}\phi^{M|c}
-2\phi^{J|a}\phi^{K|b}\phi^{L|c}\phi^{M |d}
\right) \notag \\
& ~~
+d_{IJKLM}\left(
\phi^{J |a}\phi^{K|c} \varepsilon^{bdef}  \phi^{L}_{|e}\phi^M_{|f}
+\phi^{J |a}\phi^{K|d}\varepsilon^{bcef}   \phi^L_{|e}\phi^M_{|f}
\right),
\label{xiIabcd}
\end{align}
where $a_I$,
$b_{IJK}$, $c_{IJKLM}$, and $d_{IJKLM}$ are arbitrary functions of $\phi^I$ and $X^{JK}$ satisfying
\begin{align}
&b_{IJK}=b_{IKJ},\notag \\
&c_{IJKLM}=c_{IKJLM}=c_{IJKML}=c_{ILMJK}, \notag \\
&d_{IJKLM}=-d_{IKJLM}=-d_{IJKML}=d_{ILMJK}. 
\end{align}
The explicit forms of
$\xi^{abcdef}, \xi^{abcdef}_{IJ}, \xi^{abcdefgh}$, and
$\xi^{abcdefgh}_I$ for the bi-scalar case are given in
Appendix~\ref{App:xi}.  Substituting all the $\xi$ tensors into
Eq.~(\ref{preeom2}) and rearranging the equation, we arrive at the most
general second-order rank-2 tensor $\tcalG{}^{ab}$ whose divergence is
also of second order,
\begin{align}
\tcalG{}^{a}_{b}&=A\delta^{a}_{b}+
B_{IJ}\phi^{I|a} \phi^J_{|b}+
C_I\delta^{ac}_{bd}\phi^{I|d}_{|c}+
D_{IJK}\delta^{ace}_{bdf}\phi^{I}_{|c} \phi^{J|d}\phi^{K|f}_{|e}
+
E_{IJKLM}\delta^{aceg}_{bdfh}\phi^{I}_{|c} \phi^{J|d}\phi^{K}_{|e} \phi^{L|f}\phi^{M|h}_{|g}
\notag \\  &\quad
+
F_{IJKLM}\delta^{aceg}_{bdfh}\left( 
\varepsilon_{cepq} \phi^{I|p} \phi^{J|q}\phi^{K|d} \phi^{L|f}+\phi^{I}_{|c}\phi^{J}_{|e}
\varepsilon^{dfpq}\phi^{K}_{|p}\phi^L_{|q}\right) \phi^{M|h}_{|g}
+
G_{IJ}\delta^{ace}_{bdf}\phi^{I|d}_{|c} \phi^{J|f}_{|e}
\notag \\  &\quad
+
H_{IJKL}\delta^{aceg}_{bdfh}
\phi^{I}_{|c} \phi^{J|d}\phi^{K|f}_{|e}\phi^{L|h}_{|g}+I\delta^{ace}_{bdf}R_{ce}^{~~\,df}
+
J_{IJ}\delta^{aceg}_{bdfh}\phi^{I}_{|c}\phi^{J|d}R_{eg}^{~~\,fh}+
K_{I}\delta^{aceg}_{bdfh}\phi^{I|d}_{|c}R_{eg}^{~~\,fh}\notag \\
&\quad+
L_{IJK}\delta^{aceg}_{bdfh}\phi^{I|d}_{|c} \phi^{J|f}_{|e}\phi^{K|h}_{|g}, \label{preeom3}
\end{align}
where $A, B_{IJ}, C_{I}, D_{IJK}, E_{IJKLM}, F_{IJKLM}, G_{IJ}, H_{IJKL}, I, J_{IJ}, K_{I}$,
and $L_{IJK}$ are arbitrary functions of $\phi^I$ and $X^{IJ}$, and they are subject to
\begin{align}
& B_{IJ} = B_{JI}, &
& D_{IJK}=D_{JIK}, &
& E_{IJKLM} =-E_{KJILM}=-E_{ILKJM} =E_{JILKM},
\notag \\
& G_{IJ}=G_{JI}, &
& H_{IJKL}=H_{JIKL}=H_{IJLK}, &
& F_{IJKLM} =-F_{JIKLM}=-F_{IJLKM} =F_{JILKM},
\notag \\
& J_{IJ} = J_{JI}, &
& L_{IJK} =L_{JIK}=L_{IKJ}.
\end{align}

\subsection{Consequence of Eq.~(\ref{G-Bianchi})}

In the previous subsection
we have obtained the most general second-order rank-2 tensor
whose divergence remains of second order, $\tcalG{}^{ab}$.
As can be seen, $\tcalG{}^{ab}$ involves many arbitrary functions.
It turns out, however, that those functions are not completely independent
in order for the equations of motion to be compatible with general covariance.
In this subsection,
we impose the identity~(\ref{G-Bianchi}) that arises due to general covariance,
i.e., we require that
the divergence of $\tcalG{}^{ab}$ is written as a product of $\phi^{I |a}$ and some scalar function.
This procedure will reduce the number of the arbitrary functions.
A straightforward calculation shows that 
\begin{align}
\nabla^b\tcalG{}_b^a
&=
\mathcal{Q}_I\phi^{I |a}+
 \alpha_{IJ} \delta^{ace}_{bdf}\phi^{I|d}_{|c} \phi^{J|l}R_{el}^{~~\,bf} 
+
\beta_{IJ}\delta^{ace}_{bdf}\phi^I_{|l}\phi^{J|lb}R_{ce}^{~~\,df} +
\gamma_{IJKL}\delta^{ace}_{bdf}\phi^K_{|l}\phi^{L|lb}\phi^I_{|c}\phi^{J |m}R_{em}^{~~\,df} \notag \\
&\quad+
\epsilon_{IJK} \delta^{acel}_{dfhm} \phi^{K|d}_{|c}\phi^{I|f}_{|e} \phi^{J |g}R_{gl}^{~~\,hm}
+ 
\mu_{I} \delta^{ac}_{bd}R_{cl}^{~~\,bd}\phi^{I |l} 
+
\nu_{IJKL} \delta^{ace}_{bdf} \phi^K_{|l}\phi^{L |lb}\phi^{I|d}_{|c} \phi^{J|f}_{|e} 
\notag \\ &\quad
+
\omega_{IJK}\delta^{ac}_{bd} \phi^J_{|l}\phi^{K|lb}\phi^{I|d}_{|c} 
+
\xi_{IJ}\phi^{I|l}\phi^{J|a}_{|l}
+
\zeta_{I[JK]}
\delta^{ac}_{bd}\phi^I_{|c}\phi^{J |l}\phi^{K|m}R_{lm}^{~~\,bd}
+
\iota_{IJK}\delta^{aceg}_{bdfh}\phi^J_{|l}\phi^{K|lb}\phi^{I|d}_{|c} R_{eg}^{~~\,fh}
\notag \\&\quad
+
2\eta_{I[J|K|L]}
 \delta^{ace}_{bfh}\phi^I_{|c} \phi^{J |g}\phi^{K|f}_{|e}\phi^{L|l}R_{gl}^{~~\,bh} 
+
\left( \lambda_{IJKLM}-\lambda_{ILMJK}\right) 
\delta^{ac}_{bd}\phi^L_{|e}\phi^{M |eb}\phi^I_{|c}\phi^{J |f}\phi^{K|d}_{|f}
\notag \\&\quad
+
\sigma_{IJKLMN}\delta^{acg}_{bfh}
\phi^{M}_{|l}\phi^{N |lb}\phi^I_{|c}\phi^{J|e}\phi^{K|f}_{|e} \phi^{L|h}_{|g}
+\frac{3}{2}
\tau_{IJKLM}\delta^{aceg}_{bdfh}\phi^{L|l}\phi^{M|b}_{|l} \phi^{I|d}_{|c}
\phi^{J|f}_{|e} \phi^{K|h}_{|g}
\notag \\&\quad 
+ 2 F_{IJKLM} \left(
  \varepsilon_{bdfh} \phi^{I |a} \phi^{J|g} \phi^{K|d} \phi^{L|f}
+ \varepsilon^{aceg} \phi^{I}_{|c} \phi^{J}_{|e} \phi^{K}_{|b} \phi^{L}_{|h}
\right) \phi^{M |l} R_{gl}^{~~\,bh}
\notag \\&\quad 
+ 4 \varepsilon_{bdfh} \left( F_{IJKLM} \phi^{I|a} \phi^{J |g} \right)^{|b} \phi^{K |d} \phi^{L |f} \phi^{M |h}_{|g}
+ 4 \varepsilon^{aceg} \left( F_{IJKLM} \phi^{I}_{|c} \phi^{J}_{|e}  \phi^{K}_{|b}
 \phi^{L}_{|h} \right)^{|b} \phi^{M |h}_{|g},
\label{DEOM}
\end{align}
where the coefficients are functions of $\phi^I$ and $X^{JK}$ and defined as
{\allowdisplaybreaks
\begin{align}
\alpha_{IJ} =& G_{IJ}-2J_{IJ}+2K_{I,J}-2H_{KLIJ}X^{KL} ,
\notag \\
\beta_{IJ}=&
-I_{,IJ}+J_{IJ}-K_{J,I} + 2 J_{KL,IJ}X^{KL},
\notag \\
\gamma_{IJKL}=&
-2J_{IJ,KL}+H_{IKJL},
\notag \\
\epsilon_{IJK}=& K_{(I,K)J} -\frac{3}{2}L_{KIJ},
\notag \\
\mu_{I}=& \frac{1}{2}C_I +2I_{,I} -\left( D_{JKI}+8J_{J[K,I]}\right)X^{JK}
+4E_{JKLMI}X^{JK}X^{LM},
\notag \\
 \nu_{IJKL} =&
-G_{IJ,KL} + 3 H_{K(IJL)} + 2H_{MNIJ,KL}X^{MN} - 3L_{LIJ,K},
\notag \\
\omega_{IJK} =&
-C_{I,JK} + 2 D_{J(IK)} - 2 G_{IK,J} +2\left( D_{LMI,JK} -4 H_{[J|LIK,|M]}\right) X^{LM}
\notag \\
&
-16
E_{(I|JLM|K)} X^{LM}  
-8
E_{LMNOI,JK} X^{LM} X^{NO},
\notag \\
\xi_{IJ}=&
-A_{,IJ} +B_{IJ} - C_{J,I}
-4 D_{K[I|J,|L]} X^{KL} \notag \\
&-8 E_{KLMNJ,I}X^{KL}X^{MN} +16E_{KIMNJ,L}X^{KL} X^{MN},
\notag \\
\zeta_{IJK} =&
-\frac{1}{2} D_{IJK} -2 J_{IJ,K} + 4 E_{LMIJK} X^{LM},
\notag \\
\eta_{IJKL} =&\frac{1}{2} H_{IJKL} ,
\notag \\
\lambda_{IJKLM} =&
\frac12 D_{IJK,LM} + H_{IJKM,L} - 2E_{MLIJK} - 4 E_{NOIJK,LM} X^{NO},
\notag \\
\sigma_{IJKLMN} =&
H_{IJKL,MN} - H_{IMNL,JK},
\notag \\
\tau_{IJKLM} =&
- L_{[I|JK,L|M]},
\notag \\
\iota_{IJK}=&
- K_{[I,K]J}.
\label{tau-iota}
\end{align}
}%
We present the explicit form of ${\cal Q}_I$ in Appendix~\ref{App:Q_I},
though it is irrelevant to the following derivations.

In order for the right-hand side of Eq.~(\ref{DEOM}) to be proportional
to $\phi^{I |a}$, 
all the terms that are not parallel
to $\phi^{I|a}$ must vanish identically.
We now derive the conditions for this
following the procedure of Ref.~\cite{Horndeski:1974wa}.  Let us first
focus on the $\epsilon_{IJK}$ and $\iota_{IJK}$ terms in
Eq.~(\ref{DEOM}), which are proportional to $\phi^{I|a}_{|b}\phi^{J|c}_{|d}R_{fh}^{~~\,eg}$:
\begin{equation}
\nabla^b \tcalG{}_b^a
\supset
\epsilon_{IJK}\delta^{acel}_{dfhk}
\phi^{K}{}_{|c}^{|d}
\phi^{I}{}_{|e}^{|f}
\phi^{J}{}^{|g}
R_{gl}^{~~\,hk}
+
\iota_{IJK} \delta^{aceg}_{bdfh}
\phi^J{}_{|l}
\phi^K{}^{|lb}
\phi^I{}_{|c}^{|d}
R_{eg}^{~~\,fh}
+ \cdots.
\label{epsilon-iota}
\end{equation}
The coefficient of $\phi^A{}_{,mn}\phi^B{}_{,op}g_{qr,st}$ in this
quantity can be extracted by taking a derivative with respect to
$\phi^A{}_{,mn}\phi^B{}_{,op}g_{qr,st}$ as%
\footnote{ To obtain this
result, we use the fact that derivatives of $\phi_{I,ab} \phi_{J,cd}$
and $R_{abcd}$ are given by
\begin{align*}
\frac{\partial\left( \phi^I{}_{|ab} \phi^J{}_{|cd} \right)}
{
\partial \phi^A{}_{,mn}
\partial \phi^B{}_{,op}
}
&=
\delta^I_A \delta^m_{(a}\delta^n_{b)}
\delta^J_B \delta^o_{(c}\delta^p_{d)}
+
\delta^J_A \delta^m_{(c}\delta^n_{d)}
\delta^I_B \delta^o_{(a}\delta^p_{b)}
=
\frac12\left(
\delta^I_A \delta^J_B D^{mnop}_{acbd}
+
\delta^J_A \delta^I_B D^{mnop}_{cadb}
\right),
\\
\frac{\partial R_{abcd}}{\partial g_{qr,st}}
&=
\frac14\left(
  D^{qrst}_{abcd}
+ D^{qrst}_{cdab}
- D^{qrst}_{abdc}
- D^{qrst}_{bacd}
\right),
\end{align*}
where $D^{ijkl}_{abcd}\equiv 2 \delta^i_{(a} \delta^j_{d)} \delta^k_{(b} \delta^l_{c)}$.
}
\begin{equation}
\begin{aligned}
\left(
\nabla^b \tcalG{}_b^a
\right)
\!\vphantom{)}^{;mn}_{\,A} \vphantom{)}^{;op}_{\,B} \vphantom{)}^{;qr,st}
&=
\left(
\epsilon_{IJK}
\delta^{acel}_{dfhk}
\phi^{K|d}_{|c}
\phi^{I}_{|e}{}^{|f}
\phi^{J |g}
R_{gl}^{~~\,hk}
+
\iota_{IJK} \delta^{aceg}_{bdfh}
\phi^{J}_{|l}
\phi^{K |lb}
\phi^{I|d}_{|c}
R_{eg}^{~~\,fh}
\right)
\!\vphantom{)}^{;mn}_{\,A} \vphantom{)}^{;op}_{\,B} \vphantom{)}^{;qr,st}
\\
&=
2 \epsilon_{(A|J|B)}
\delta^{acel}_{dfhk}
\delta^{( m}_{c}g^{n)d}
\delta^{(o}_eg^{p)f}
\left(
\phi^{J|(q}g^{r)k}\delta_l^{(s}g^{t)h}
+ g^{h(q}\delta^{r)}_l \phi^{J|(s} g^{t)k} 
\right)
\\
&\quad
+ 2\iota_{IJK} \delta^{aceg}_{bdfh} \left(  
\delta^K_A \delta^I_B \phi^{J| (m}g^{n)b}\delta_c^{(o}g^{p)d}
+ \delta^I_A \delta^K_B \phi^{J| (o}g^{p)b}\delta_c^{(m}g^{n)d}
\right) \delta_e^{(q}g^{r)h} \delta_g^{(s}g^{t)f}
.
\label{epsilon-iota_diff}
\end{aligned}
\end{equation}
Equation~(\ref{epsilon-iota_diff}) must vanish when contracted with a vector $Y_a$ such that $Y_a \phi^{I|a}=0$
because Eq.~(\ref{G-Bianchi}) implies that
\begin{equation}
Y_a \left(\nabla^b \tcalG{}_b^a\right)\!\vphantom{)}^{;mn}_{\,A} \vphantom{)}^{;op}_{\,B} \vphantom{)}^{;qr,st}
= 
Y_a \Bigl(\frac12 \mathcal{E}_I \phi^{I |a}\Bigr)\!\vphantom{)}^{;mn}_{\,A} \vphantom{)}^{;op}_{\,B} \vphantom{)}^{;qr,st}
=
\frac12 Y_a \phi^{I |a} \mathcal{E}_I \vphantom{)}^{;mn}_{\,A} \vphantom{)}^{;op}_{\,B} \vphantom{)}^{;qr,st}
= 0.
\end{equation}
Thus, we obtain a constraint equation given by
\begin{multline}
2 \epsilon_{(A|J|B)}
Y_a
\delta^{acel}_{dfhk}
\delta^{( m}_{c}g^{n)d}
\delta^{(o}_eg^{p)f}
\left(
\phi^{J| (q}g^{r)k}\delta_l^{(s}g^{t)h}
+ g^{h(q}\delta^{r)}_l \phi^{J|(s} g^{t)k}
\right)
\\
+ 2\iota_{IJK} 
Y_a
\delta^{aceg}_{bdfh} \left(  
\delta_A^K \delta_B^I \phi^{J| (m}g^{n)b}\delta_c^{(o}g^{p)d}
+ \delta_A^I \delta_B^K \phi^{J| (o}g^{p)b}\delta_c^{(m}g^{n)d}
\right) \delta_e^{(q}g^{r)h} \delta_g^{(s}g^{t)f}
=0.
\label{epsilon-iota_constraint}
\end{multline}
This constraint has eight free indices $m,n,o,p,q,r,s,t$, and its any
component must be fulfilled.  We first take the trace of
Eq.~(\ref{epsilon-iota_constraint}) by contracting with
$g_{mn}g_{op}g_{qr}$, giving
\begin{equation}
-8 \epsilon_{(A|J|B)} \phi^{J (s} Y^{t)}
+ 8 \iota_{(A|J|B)} \phi^{J (s} Y^{t)}
=
-8 \epsilon_{(A|J|B)} \phi^{J (s} Y^{t)}
=0,
\end{equation}
where we have used $\iota_{(A|J|B)}=0$ which follows from the
definition~(\ref{tau-iota}).  This equation must be satisfied for any
$\phi^{J|a}$, and therefore it is necessary to impose
$\epsilon_{(A|J|B)}\left(=\epsilon_{AJB}\right) = 0$.
Further constraints can be derived from Eq.~(\ref{epsilon-iota_constraint}) as follows.
Let us project Eq.~(\ref{epsilon-iota_constraint}) to the basis vectors
$Y^a,\tilde Y^a,\phi^{I|a}$ ($I=1,2$) that satisfy
\begin{equation}
Y_a Y^a = \tilde Y_a \tilde Y^a = 1,
\qquad
Y_a \tilde Y^a = 0,
\qquad
Y_a \phi^{I|a} = \tilde Y_a \phi^{I|a} = 0\;\;\;{\rm for}\;\;\; I=1,2
.
\end{equation}
By contracting Eq.~(\ref{epsilon-iota_constraint}) with $Z_m
W_n V_o \phi^{C}_{|p}\phi^{D}_{|q}\phi^{E}_{|r}\phi^{F}_{|s}\phi^{G}_{|t}$, where $Z^a,
W^a,$ and $V^a$ are either $Y^a$ or $\tilde Y^a$, we find
\begin{align}
0&=
-4\epsilon_{(A|J|B)}
\delta^{ac}_{bd}Y_a V^b 
\bigl(Z_c W^d + W_c Z^d \bigr)
\left(
2  X^{(J|(D}X^{E)(F}X^{G)|C)} 
- X^{J(D}X^{E)C}X^{FG}
- X^{DE}X^{C(F}X^{G)J}
\right)
\notag \\ &\quad
+4
\iota_{AJB} 
\delta^{ac}_{bd}Y_a V^b 
\bigl(Z_c W^d + W_c Z^d \bigr)
X^{JC}\left(
X^{(D|(F}X^{G)|E)} - X^{DE}X^{FG}
\right).
\label{epsilon-iota2}
\end{align}
Using $\epsilon_{(A|J|B)}=0$
obtained in the previous step, we find $\iota_{AJB}=0$ as another constraint to be imposed.

Repeating a similar procedure for any other products of
the second derivative terms in Eq.~(\ref{DEOM}), we find the following constraint equations for the coefficient functions:
\begin{gather}
\alpha_{IJ} = -2 \beta_{JI},
\quad
\gamma_{IJKL} = -4 \eta_{I[J|L|K]},
\quad
\alpha_{AI}\Bigl[ \delta_{I(C} X^{-1}_{\smash{D)B}} - \delta_{IB} X^{-1}_{CD} \Bigr] 
= 4 \left( \eta_{(CD)AB} -  \eta_{(C|BA|D)} \right)
,
\label{coeffcond1}
\\
 \epsilon_{IJK} = \iota_{IJK}
= \omega_{IJK} = \lambda_{(IJ)KLM} - \lambda_{(I|LM|J)K}
= \mu_I = \zeta_{I[JK]}
= \xi_{IJ} = \tau_{IJKLM}
= F_{IJKLM}
=0,
\label{coeffcond2}
\\
\nu_{ACDB} X^{-1}_{EF} - \nu_{AB(E|C} X^{-1}_{\smash{|F)D}} 
-2 \sigma_{(EF)CADB} 
= 0,
\qquad
\nu_{B[A|K|C]} = \sigma_{EF[CA]DB} = 0,
\label{coeffcond3}
\end{gather}
where $X^{-1}_{IJ}$ is the inverse matrix of $X^{IJ}$.

The constraints~(\ref{coeffcond1}) and (\ref{coeffcond2}) impose the
following conditions on the functions appearing in Eq.~(\ref{preeom3}):
{\allowdisplaybreaks
\begin{align}
B_{IJ}&=
- 2  \left( \mathcal{F} + 2 \mathcal{W} \right)_{,I,J} 
+ A_{,IJ}
+ 2 D_{(I|K|J),L} X^{KL}
- 16 E_{K(I|MN|J),L} X^{KL} X^{MN}\notag \\
&\quad- 8 \left( J_{K(I,J),L} - J_{KL,I,J} \right) X^{KL},
\label{coeff_top}
\\
C_{I}&=
-2\left(\mathcal{F} + 2\mathcal{W}\right)_{,I}
+ 2 \left( D_{JKI} + 8 J_{J[K,I]} \right) X^{JK}
- 8 E_{JKLMI} X^{JK} X^{LM}, 
\\
F_{IJKLM} &= 0,
\\
G_{IJ}
&=
2J_{IJ} - 2 K_{(I,J)} + 4 J_{K(I,J)L} X^{KL},
\\
H_{IJKL} &= 2 J_{IJ,KL},
\\
K_{[I,J]} &= -2 J_{K[I,J]L} X^{KL},
\\
K_{I,JK} &= K_{J,IK},
\\
L_{IJK} &= \frac23 K_{(I,JK)},
\\
I&=\frac{1}{2}\mathcal{F}+\mathcal{W}, 
\label{coeff_bottom}
\end{align}
}%
where
$\mathcal{W} = \mathcal{W}(\phi^I)$, and $\mathcal{F}=\mathcal{F}(\phi^I, X^{JK})$ is a function satisfying
$\mathcal{F}_{,IJ}=G_{IJ}$, which is integrated to give
\begin{equation}
\mathcal{F}
=
\int G_{IJ} \, dX^{IJ}
=
\int \left(
2 J_{IJ} - 2 K_{I,J} + 4 J_{KI,JL} X^{KL} 
\right) dX^{IJ}.
\label{Fintegral}
\end{equation} 
The conditions
$\zeta_{I[JK]} = \omega_{IJK} = \lambda_{(IJ)KLM} - \lambda_{(I|LM|J)K} =0$ in Eq.~(\ref{coeffcond2}) imply
\begin{align}
D_{I[JK]}
&=
- 4 J_{I[J,K]} + 8 E_{LMI[JK]} X^{LM},
\label{derconst_top}
\\
 D_{I(JK)}
&=
\frac12 C_{J,IK}
+ G_{JK,I}
+ \left( -  D_{LMJ,IK} + 4 H_{[I|LJK,|M]} \right) X^{LM}
\notag \\ &\quad 
+ 8 E_{(J|ILM|K)} X^{LM}
+ 4 E_{LMNOJ,IK} X^{LM} X^{NO},\label{DIJKanticond}
\\
0&=
\frac12 D_{IJK,LM} - \frac12 D_{(I|LM,|J)K}
+ H_{IJKM,L} - H_{(I|LMK,|J)}
\notag \\ & \quad
- 2\left( E_{ML(IJ)K} - E_{K(IJ)LM} \right)
- 4 \left( E_{NO(IJ)K,LM} - E_{NO(I|LM,|J)K} \right) X^{NO},
\end{align}
and Eq.~(\ref{coeffcond3}) implies
\begin{align}
G_{I[J,K]L} &= 0,
\label{GIJKLcond}
\\
H_{IJK[L,M]N} &= 0,
\\
G_{(IJ,KL)} &=  3 H_{L(IJK)} + 2 H_{LM(IJ,KN)} X^{MN} - 2 K_{(I,JK),L}
~.
\label{derconst_bottom}
\end{align}
Equation~(\ref{GIJKLcond}) is nothing but the integrability condition
which guarantees 
$\mathcal{F}_{,IJ,KL}=\mathcal{F}_{,KL,IJ}$, and hence
the integral~(\ref{Fintegral}) indeed exists.

\subsection{The most general second-order equation of motion}

Now we are at the final stage of deriving the most general second-order field
equations of the bi-scalar-tensor theory.  Substituting
Eqs.~(\ref{coeff_top})--(\ref{coeff_bottom}) into Eq.~(\ref{preeom3}),
we at last obtain
\begin{align}
\calG{}^{a}_{b}&=
A\delta^{a}_{b}
+\left[
-2 \mathcal{F}_{,I} -4 \mathcal{W}_{,I}
+2\left( D_{JKI} + 8 J_{J[K,I]}\right) X^{JK}
-8 E_{JKLMI} X^{JK} X^{LM}
\right] \delta^{ac}_{bd}\phi^{I |d}_{|c}
\notag \\
&
+\left( -2 \mathcal{F}_{,I,J} -4\mathcal{W}_{,I,J} + A_{,IJ}
+2 D_{IKJ,L} X^{KL}
-16 E_{KIMNJ,L}X^{KL}X^{MN}
-16 J_{K[I,L],J} X^{KL}
\right)\phi^{(I|a} \phi^{J)}_{|b}
\notag \\&
+ D_{IJK}\delta^{ace}_{bdf}\phi^I_{|c} \phi^{J |d}\phi^{K|f}_{|e}
+ E_{IJKLM}\delta^{aceg}_{bdfh}\phi^{I}_{|c} \phi^{J |d}\phi^K_{|e} \phi^{L|f}\phi^{M |h}_{|g}
+\left( \frac{1}{2}\mathcal{F}+\mathcal{W}\right) \delta^{ace}_{bdf}R_{ce}^{~~\,df}
+ \mathcal{F}_{,IJ} 
\delta^{ace}_{bdf}\phi^{I |d}_{|c} \phi^{J|f}_{|e}
\notag \\&
+
J_{IJ}\delta^{aceg}_{bdfh}\phi^I_{|c}\phi^{J |d}R_{eg}^{~~\,fh}
+ 2 J_{IJ,KL}\delta^{aceg}_{bdfh}\phi^I_{|c} \phi^{J|d}\phi^{K |f}_{|e}\phi^{L|h}_{|g}
+K_{I}\delta^{aceg}_{bdfh}\phi^{I|d}_{|c}R_{eg}^{~~\,fh}
+\frac23 K_{I,JK}
\delta^{aceg}_{bdfh}\phi^{I|d}_{|c} \phi^{J|f}_{|e}\phi^{K|h}_{|g} .
\label{EoM}
\end{align}
This is the main result of this paper. The most general
field equations for the single-scalar case~\cite{Horndeski:1974wa} are
reproduced as should be if one restricts the number of the scalar fields in
Eq.~(\ref{EoM}) to one.  Note that one can eliminate ${\cal
W}(\phi^I)$ from the above equation by redefining ${\cal F}\to\hat{\cal
F}(\phi^I, X^{JK})={\cal F}+2{\cal W}$.  We can see that
Eqs.~(\ref{DIJKanticond})--(\ref{derconst_bottom}) do not reduce the
number of the arbitrary functions because these are the relations
between derivatives of the functions.  In other words, these do not
affect the structure of the field equations~(\ref{EoM}).  As we will
comment in the final section, these may, however, help us to check the
integrability conditions for the field equations.

From the relation (\ref{G-Bianchi}), the scalar-field equations of motion are found to be
\begin{align}
{\cal E}_I = 2 {\cal Q}_I + \delta^{cegl}_{bdhm} \left(
             - \gamma_{JIKL} \phi^{K|b} \phi^{J}_{|c} \phi^{L|d}_{|e} 
             R_{gl}^{~~\,hm} 
             + \frac23 \sigma_{JIKLMN} \phi^{J}_{|c} \phi^{M|b}
             \phi^{K|d}_{|e} \phi^{L|h}_{|g} \phi^{N|m}_{|l}
             \right).
\end{align}

\section{Comparison with the generalized multi-Galileon theory}
\label{Sec:multiG}


The covariant version of the multi-Galileon theory in the flat
spacetime~\cite{Padilla:2012dx} was conjectured to be the most
general multi-scalar-tensor theory with second-order field equations.
However, later it was pointed out that this theory is not the
most general one~\cite{Kobayashi:2013ina}.  A counter-example is given
by the multi-field DBI Galileons~\cite{RenauxPetel:2011uk}.  In this
section, we compare the most general second-order field equations
obtained in the previous section with the field equations of the
generalized multi-Galileons, and identify the terms that are missing in
the latter theory.

The action of the generalized multi-Galileons is given
by~\cite{Padilla:2012dx}
\begin{align}
\frac{1}{\sqrt{-g}}\mathcal{L}&=G_2-G_3{}_I \phi_I{}_{|a}^{|a} 
+G_4R+G_4{}_{,IJ}\left( \phi^{I|a}_{|a} \phi^{J|b}_{|b} -\phi^I_{|ab} \phi^{J|ab}\right) \notag \\
                  &\quad
+G_{5I}G^{ab}\phi^I_{|ab}-\frac{1}{6}G_{5I,JK}\left( 
\phi^{I|a}_{|a}\phi^{J|b}_{|b}\phi^{K|c}_{|c}-3 \phi^{I|a}_{|a}\phi^{J|c}_{|b}\phi^{K|b}_{|c}
+2 \phi^{I|b}_{|a} \phi^{J|c}_{|b} \phi^{K|a}_{|c} \right),
\end{align} 
where $G_2, G_3{}_I, G_4$ and $G_5{}_I$ are arbitrary functions of
$\phi^I$ and $X^{IJ}$, and $G^{ab}$ is the Einstein tensor.  The
functions $G_3{}_{I}{}_{,JK}, G_4{}_{,IJ,KL}, G_5{}_{I,JK}$ and
$G_5{}_{I,JK,LM}$ are totally symmetric with respect to all of their
indices, $I,J,K,L$ and $M$ in order for the field equations to be of
second order.  A straightforward calculation leads to the field
equations for the generalized multi-Galileons,
\begin{align}
E^{ab}(\mathcal{L})&=
\left( -\frac{1}{2}G_2+G_{3(I,J)}X^{IJ}-2G_{4,I,J}X^{IJ}\right) g^{ab}
+\left( -\frac{1}{2}G_{2,IJ}+G_{3(I,J)}-G_{4,I,J}\right) \phi^{I|a}\phi^{J|b}
\notag \\ & \quad
+\left( -X^{JK}G_{3IJK}+G_{4,I}+2X^{JK}G_{4IJ,K}\right) g^{l(a}\delta^{b)c}_{ld}\phi^{I|d}_{|c}
\notag \\ & \quad
+\left( -\frac{1}{2}G_{3IJK}+2G_{4K(I,J)}-\frac{1}{2}G_{5K,I,J}\right)
g^{l(a}\delta^{b)ce}_{ldf}\phi^{I}_{|c}\phi^{J|d}\phi^{K|f}_{|e}
\notag \\
&\quad -\frac14 \left( G_4 - 2 G_{4IJ}X^{IJ}
 + G_{5(I,J)}X^{IJ}\right)g^{l(a}\delta^{b)ce}_{ldf}R_{ce}^{~~\,df} \notag \\
& \quad+\left( \frac{1}{2}G_{4IJ}+X^{KL}G_{4IJKL}-\frac{1}{2}G_{5(I,J)}-\frac{1}{2}X^{KL}G_{5IJK,L}\right)
 g^{l(a}\delta^{b)ce}_{ldf}\phi^{I|d}_{|c}\phi^{J|f}_{|e}
\notag \\
& \quad
+\frac{1}{4}\left( G_{4IJ}-G_{5(I,J)}\right) g^{l(a}\delta^{b)ceg}_{ldfh}
 \phi^I_{|c} \phi^{J |d}R_{eg}^{~~\,fh}
\notag \\
& \quad+\frac{1}{2}\left(G_{4IJKL}-G_{5KL(I,J)} \right)
g^{l(a}\delta^{b)ceg}_{ldfh}\phi^I_{|c}\phi^{J|d}\phi^{K|f}_{|e} \phi^{L|h}_{|g}
\notag \\
& \quad
-\frac{1}{4}X^{JK}G_{5IJK} g^{l(a}\delta^{b)ceg}_{ldfh} \phi^{I|d}_{|c} R_{eg}^{~~\,fh}
-\frac{1}{6}\left( G_{5IJK}+X^{LM}G_{5IJKLM}\right)
g^{l(a} \delta^{b)ceg}_{ldfh} \phi^{I|d}_{|c} \phi^{J|f}_{|e} \phi^{K|h}_{|g}.
\label{EoMmultiG}
\end{align}
Comparing Eq.~(\ref{EoMmultiG}) with Eq.~(\ref{EoM}), it is easy to see
the exact correspondence between each term.  It is also found that
terms corresponding to $E_{IJKLM}$ are lacking in the generalized
multi-Galileon; this is a completely new term.  We would, however,
emphasize that, even setting $E_{IJKLM}=0$, Eq.~(\ref{EoM}) covers a
wider class of theories than the generalized multi-Galileons.  This fact
is to be illustrated in a concrete example presented below. Note in
passing that the coefficient functions of the above equation satisfy all
the constraints~(\ref{coeffcond1})--(\ref{coeffcond3}) found in the
previous section.


The double-dual Riemann term deduced from the multi-field DBI Galileons,
\begin{align}
\mathcal{L}=
\sqrt{-g} \, 
\delta_{IJ} \delta_{KL}\delta^{aceg}_{bdfh} \phi^I_{|a}\phi^{J |b} 
 \phi^K_{|c}\phi^{L|d}  R_{eg}^{~~\,fh},
\label{LddRieman}
\end{align}
is not included in the Lagrangian of the generalized multi-Galileon
theory~\cite{Kobayashi:2013ina}.  One can however check that this term
is actually contained in our theory.  It is straightforward to derive
the field equations from Eq.~(\ref{LddRieman}):
\begin{align}
E^{ab}(\mathcal{L})
=
4g^{l(a}\delta^{b)ceg}_{ldfh} X^{I[I}\phi^{J]}_{|c}\phi^{J|d}R_{eg}^{~~\,fh}
+8 g^{l(a}\delta^{b)ceg}_{ldfh} \delta_{I[J} \delta_{K]L}
\phi^I_{|c}\phi^{J|d} \phi^{K|f}_{|e} \phi^{L|h}_{|g}
~.
 \label{DDR}
\end{align}
This is reproduced by setting
$J_{IJ} = 2 \left(\delta_{IJ}\delta_{KL}-\delta_{IK}\delta_{JL}\right)X^{KL} $ in our field equations.

\section{Candidate Lagrangian}
\label{Sec:L}

Having thus determined the most general second-order field equations of
the bi-scalar-tensor theory, let us now explore the Lagrangian that
gives the field equations we have derived.  For the construction of the
Lagrangian, we employ the same strategy as taken in
Ref.~\cite{Horndeski:1974wa}.  In the single scalar-field case,
Horndeski found that the form of the Lagrangian can be guessed from the
trace of the gravitational field equation.  In the same way as in the
single-field theory, we take the trace of the field equations of the
bi-scalar-tensor theory~(\ref{EoM}) and arrive at the terms of the
following form as a candidate Lagrangian:
\begin{align}
\mathcal{L}_1&=
 \sqrt{-g} \, 
M^{(1)}{}_I\phi^{I|c}_{|c},\label{LagM1}\\
\mathcal{L}_2&= 
\sqrt{-g} 
\left( M^{(2)}{}\delta^{ce}_{df}R_{ce}^{~~\,df}
+2 M^{(2)}_{,IJ}
\delta^{ce}_{df}\phi^{I|d}_{|c} \phi^{J|f}_{|e}
\right),
\label{LagM2}\\
\mathcal{L}_3&=
\sqrt{-g} \, M^{(3)}_{IJK}\delta^{ce}_{df}\phi^{I}_{|c}\phi^{J|d}\phi^{K|f}_{|e},
\label{LagM3}\\
\mathcal{L}_4&=
\sqrt{-g} 
\left( 
M^{(4)}_{I}\delta^{ceg}_{dfh}\phi_I{}_{|c}^{|d}R_{eg}^{~~\,fh}
+\frac{2}{3} M^{(4)}_{I,JK} \delta^{ceg}_{dfh}\phi^{I|d}_{|c} \phi^{J|f}_{|e} \phi^{K|h}_{|g}
\right),
\label{LagM4}\\
\mathcal{L}_5&=
\sqrt{-g} 
\left( 
M^{(5)}_{IJ}\delta^{ceg}_{dfh}\phi^I_{|c}\phi^{J|d}R_{eg}^{~~\,fh}
+2 M^{(5)}_{IJ,KL} \delta^{ceg}_{dfh}\phi^{I}_{|c}\phi^{J|d}\phi^{K|f}_{|e}\phi^{L|h}_{|g}
\right), 
\label{LagM5}\\
\mathcal{L}_{6}&=\sqrt{-g} \, M^{(6)}, \label{LagM6}\\
\mathcal{L}_7&= 
\sqrt{-g} \, 
M^{(7)}_{IJKLM}\delta^{ceg}_{dfh}\phi^I_{|c}\phi^{J|d}\phi^K_{|e}\phi^{L|f}\phi^{M|h}_{|g}
,
\label{LagM7}
\end{align} 
where $M^{(1)}, M^{(2)}, M^{(3)}_{IJK}, M^{(4)}_I, M^{(5)}_{IJ},
M^{(6)}$, and $M^{(7)}_{IJKLM}$ are arbitrary functions of $\phi^I$ and
$X^{IJ}$ satisfying
\begin{align}
&M^{(3)}_{IJK}=M^{(3)}_{JIK},\\
&M^{(5)}_{IJ}=M^{(5)}_{JI},\\
&
M^{(7)}_{IJKLM}=-M^{(7)}_{KJILM}=-M^{(7)}_{ILKJM}=M^{(7)}_{JILKM}.
\end{align}
In order to maintain the second-order equations of motion for the scalar fields,
we have to impose extra conditions on these functions.
For example, the Euler-Lagrange equation of $\mathcal{L}_1$ for the scalar field $\phi^I$ is given by
\begin{align}
E_I\left( \mathcal{L}_{1}\right) &=M^{(1)}_{J,KI}
\phi^{J|cd}_{|c}\phi^K_{|d}-M^{(1)}_{I,JK}\phi^{J|dc}_{|c}\phi^K_{|d}
-2M^{(1)}_{J,KI,L}X^{KL}\phi^{J|c}_{|c}
+2M^{(1)}_{I,JK,L}X^{JK}_{|c}\phi^{L|c}
\notag \\ &\quad 
+M^{(1)}_{J,KI,LM}X^{LM|d}\phi^K_{|d}\phi^{J|c}_{|c}
+M^{(1)}_{I,JK,LM}X^{JK}_{|c}X^{LM|c}
\notag \\&\quad 
+
2 M^{(1)}_{(I,J)}
\phi^{J|c}_{|c}
-2M^{(1)}_{I,J,K}X^{JK}
+M^{(1)}_{J,KI}\phi^{J|c}_{|c}\phi^{K|d}_{|d}
-M^{(1)}_{I,JK}\phi^{J|d}_{|c}\phi^{K|c}_{|d}.\label{scalareq1}
\end{align}
The two terms in the first line of Eq.~(\ref{scalareq1}) are of third order,
while the other terms are of second or first order. 
To eliminate the third-order derivatives, we therefore impose
\begin{align}
M^{(1)}_{[I,J]K}=0.
\end{align}
Performing the same analysis,
we find that higher-derivative terms are removed by requiring that
\begin{equation}
\begin{aligned}
&M^{(2)}_{,I[J,K]L}=0,
\qquad
M^{(3)}_{IJ[K,L]M}=0, \qquad
M^{(4)}_{[I,J]K}=0,\qquad
M^{(4)}_{I,J[K,L]M}=0,\qquad
\\
&M^{(5)}_{IJ,K[L,M]N}=0,\qquad
M^{(7)}_{IJKL[M,N]O}=0. 
\end{aligned}
\label{condi2ndScalar}
\end{equation}

The Euler-Lagrange equations
for the Lagrangian densities (\ref{LagM1})--(\ref{LagM7}) are listed in Appendix~\ref{App:ELeq},
from which
one can see the relations between the functions appearing in Eq.~(\ref{EoM}) and
those in Eqs.~(\ref{LagM1})--(\ref{LagM7}):
\begin{align}
A&=
 M^{(1)}_{I,J} X^{IJ}
+ 4 M^{(2)}_{,I,J}X^{IJ}
- 2 M^{(3)}_{IJK,L} X^{IJ} X^{KL}
-8 \left( M^{(5)}_{IJ,K,L} - M^{(5)}_{IL,J,K} \right) X^{IJ} X^{KL} 
\notag \\ & \quad
+\frac{1}{2}M^{(6)}
+ 8 M^{(7)}_{IJKLM,N} X^{IJ} X^{KL} X^{MN},
\\
D_{IJK}&=
-\frac12 M^{(1)}_{(I,J)K}
-4 M^{(2)}_{,(I,J)K}
+\frac32 M^{(3)}_{(IJK)}
+ \left( M^{(3)}_{L(IJ),MK} +M^{(3)}_{(I|LM,|J)K} -M^{(3)}_{IJL,KM} \right) X^{LM}
\notag \\&\quad
-2 M^{(4)}_{K,I,J}
+2\left( 2 M^{(5)}_{K(I,J)} - M^{(5)}_{IJ,K} \right)
+ 4 \left( 2M^{(5)}_{L(I,J),MK} - M^{(5)}_{IJ,L,MK} \right) X^{LM}
\notag \\
&\quad
+ 12 M^{(7)}_{IJ(KLM)} X^{LM}
+ 8 M^{(7)}_{IJNLM,OK} X^{LM} X^{NO}, 
\\
 E_{IJKLM} 
&=
\frac{1}{8}\left( \delta_{IK}^{PQ}\delta_{JL}^{RS}+\delta_{JL}^{PQ}\delta_{IK}^{RS}\right) 
\biggl[
-\frac12 M^{(3)}_{PRQ,SM}
-4 M^{(5)}_{PR,Q,SM}
+M^{(7)}_{PRQSM}+M^{(7)}_{PRMSQ}+M^{(7)}_{PRQMS}
\notag \\ & 
\qquad\qquad\qquad\qquad\qquad\quad
-\left(
M^{(7)}_{PRQSN,OM} - 2 M^{(7)}_{PRNSQ,OM} - 2 M^{(7)}_{PNQSO,RM}
\right)X^{NO}
\biggr],
\\
\mathcal{F}+2\mathcal{W}&=
M^{(2)}
-2 M^{(2)}_{,IJ}X^{IJ}
-2 M^{(4)}_{I,J}X^{IJ}
+2M^{(5)}_{IJ}X^{IJ}
+4 M^{(5)}_{IK,JL}X^{IJ}X^{KL},
\\
J_{IJ}&=
-\frac12 M^{(2)}_{,IJ}
- M^{(4)}_{(I,J)}
+M^{(5)}_{IJ}
+ \left( 2M^{(5)}_{K(I,J)L} - M^{(5)}_{IJ,KL}  \right) X^{KL},
\\
K_I&=
-M^{(4)}_{J,KI}X^{JK}. \label{LangK}
\end{align}
In addition, comparing the $\phi^{(I}{}^{|a} \phi^{J)}{}_{|b}$ and
$\delta^{ac}_{bd}\phi^I{}^{|d}_{|c}$ terms, we see that the
following two conditions must be satisfied:
\begin{align}
&
-2 
\left(\mathcal{F}+2\mathcal{W}\right)_{,I,J}
+ A_{,IJ}
+ 2 D_{K(IJ),L} X^{KL}
+16 E_{KMN(IJ),L} X^{KL} X^{MN}
-8\left( J_{K(I,J),L} - J_{KL,I,J} \right)X^{KL}
\notag \\
&=
M^{(1)}_{(I,J)}
+ 2 M^{(2)}_{,I,J}
-\left( M^{(3)}_{IJK,L} + 2 M^{(3)}_{KL(I,J)} - 2 M^{(3)}_{K(IJ),L} \right) X^{KL}
\notag \\& \quad
- 4 \left(
M^{(5)}_{IJ,KL} + M^{(5)}_{KL,IJ} - 2 M^{(5)}_{IK,JL}
\right) X^{KL}
+\frac12 M^{(6)}_{,IJ}
\notag \\ & \quad
+ 8
\left( M^{(7)}_{MNKL(I,J)} - 2 M^{(7)}_{MN(KI)J,L} + M^{(7)}_{MN(IJ)K,L} \right) X^{KL} X^{MN},
\\
\notag \\
& -2 \left(\mathcal{F} + 2\mathcal{W} \right) _{,K}
+2 \left( D_{IJK}  + 8 J_{I[J,K]} \right) X^{IJ}
-8 E_{IJLMK} X^{IJ} X^{LM}
\notag \\ &
=
 -M^{(1)}_{I,JK} X^{IJ}
-2 \left( M^{(2)}_{,K} + 2 M^{(2)}_{I,JK}X^{IJ} \right)
+ 3 M^{(3)}_{(IJK)} X^{IJ}
+2 M^{(3)}_{IJM,LK} X^{IL} X^{JM}.
\end{align}

The Lagrangian is constructed by solving the above equations for $M^{(1)},\ldots,M^{(7)}$, 
and unfortunately we have not accomplished this step yet.
In Appendix~\ref{App:CLS}, we review the construction of the Lagrangian for the single-scalar case. 
The lesson from the single-field Lagrangian is that
we should probably integrate Eq.~(\ref{LangK}) first to
identify $M^{(4)}$ compatible with Eq.~(\ref{condi2ndScalar}).
We have not yet succeeded even in solving these equations.
In addition, it should be kept in mind that the terms we considered ($M^{(1)},\ldots,M^{(7)}$) might
not be enough to construct the true Lagrangian.
Those terms are simply inferred from the trace of the equations of motion,
and in general there is no guarantee that they are enough though they happened to be so in the single scalar-field case.
Albeit these difficulties, we believe that above
calculations are useful to build the Lagrangian for our theory. We hope
to report on this final part of the construction of the most general second-order
bi-scalar-tensor theory in the near future.

\section{Discussions and summary}
\label{Sec:Discussion}

In this paper, we have reported our attempt to construct the bi-scalar
generalization of the Horndeski theory in four-dimensional spacetime.
Following Horndeski's method, we have succeeded in deriving all the
possible terms appearing in the most general second-order field
equations for the bi-scalar tensor theory. We compared our field
equations with those of the generalized multi-Galileon theory, and identified
the terms that are really not included in that theory.
In particular, we confirmed that the double-dual Riemann term, which was
shown to be missing in that theory~\cite{Kobayashi:2013ina}, 
can be reproduced from our results by choosing
the arbitrary functions appropriately.
We have also discussed the construction of the Lagrangian yielding
the field equations we found.  For the construction, we have taken
Horndeski's approach based on the trace of the field equations as
a candidate Lagrangian.
From the Euler-Lagrange equations we have obtained several differential
equations for the functions in the Lagrangian, 
though we could not solve them to give explicit forms of the functions.
In fact, it is still unclear whether or not the candidates of the Lagrangian
we proposed suffice to generate all the terms in
our most general field equations. 

As discussed {\em e.g.}\ in Ref.~\cite{Deffayet:2013tca}, we have to
impose the integrability conditions on the field equations in order to
ensure the existence of a corresponding Lagrangian. The integrability conditions are
summarized as
\begin{align}
& \frac{ \delta \sqrt{-g} \, \calG^{\mu \nu}(x)}{\delta g_{\rho \lambda}(y)}-\frac{\delta \sqrt{-g} \, \calG^{\rho \lambda}(y)}{\delta g_{\mu \nu}(x)}=0, \\
&  \frac{\delta \sqrt{-g} \, \calG^{\mu \nu}(x)}{\delta \phi^{I}(y)}- \frac{\delta \sqrt{-g}\,  \mathcal{E}_I(y)}{\delta g_{\mu \nu}(x)}=0, \\
& \frac{\delta \sqrt{-g} \, {\cal E}_I(x)}{\delta \phi^{J}(y)}-\frac{\delta \sqrt{-g} \, \mathcal{E}_J(y)}{\delta \phi^{I}(x)}=0,
\end{align}
where $\delta/\delta A$ denotes variation with respect to a field
$A$.  Surprisingly, Horndeski found the corresponding Lagrangian in a
rather heuristic way without using the integrability conditions
explicitly. This implies that, in the single scalar case, those
conditions are automatically satisfied and do not give rise to any extra
constraints on the functions in the field equations.
It is unclear that
this property persists in theories with multiple scalar fields,
and we have not been able to determine the corresponding Lagrangian
completely. The integrability conditions might give us some clues to
accomplish this procedure. We hope to report the result obtained
from such an approach in the near future.

\acknowledgments
We would like to thank Xian Gao, Emir G\"umr\"uk\c{c}\"uo\u{g}lu,
Hideo Kodama, Shinji Mukohyama, Vishagan Sivanesan and Yi Wang for fruitful
discussions and useful comments. S.O.\ was supported by JSPS
Grant-in-Aid for Scientific Research No.\ 25-9997. N.T.\ was supported by
the European Research Council grant no.\ ERC-2011-StG 279363-HiDGR. This
work was supported in part by the JSPS Grant-in-Aid for Scientific
Research Nos.~24740161 (T.K.), 25287054 and 26610062 (M.Y.).

\appendix

\section{The $\xi$ tensors}
\label{App:xi}

In this appendix, we show the expressions for the $\xi$ tensors
introduced in Sec.~\ref{Sec:xi} to construct Eq.~(\ref{preeom3}).
Expressions of $\xi^{ab}$ and $\xi_I^{abcd}$ are given by
Eqs.~(\ref{xiab}) and (\ref{xiIabcd}), and we need to construct the
other $\xi$ tensors appearing in Eq.~(\ref{preeom2}), {\em i.e.,}
$\xi^{abcdef}$, $\xi^{abcdef}_{IJ}$, $\xi^{abcdefgh}$, and
$\xi^{abcdefgh}_I$.  As described in Sec.~\ref{Sec:xi}, the $\xi$
tensors are constructed by taking the products and linear combinations
of $\phi^I{}^{|a}$, $g^{ab}$ and $\varepsilon^{abcd}$ that enjoy
property $S$, and we need to find all such $\xi$ tensors to construct
the most general second-order tensor whose divergence remains of second order.
In doing so, we find that not all of the $\xi$ tensors give
nontrivial contributions to Eq.~(\ref{preeom2}) because there is some degeneracy and the identical
terms appear from more than one $\xi$ tensor.
Those $\xi$ tensors that give nontrivial contributions to
Eq.~(\ref{preeom2}) are summarized as
{\allowdisplaybreaks
\begin{align}
&\xi_{}^{abcdef}
=
\hat a_{IJ}\Bigl(
 \epsilon^I{}^{ace}\epsilon^J{}^{bdf}
+\epsilon^I{}^{bce}\epsilon^J{}^{adf}
+\epsilon^I{}^{ade}\epsilon^J{}^{bcf}
+\epsilon^I{}^{bde}\epsilon^J{}^{acf}
\Bigr)
\notag \\ & \qquad\quad
+ \hat b\,
g_{gh}\Bigl( \varepsilon^{aceg}\varepsilon^{bdfh}+\varepsilon^{bceg}\varepsilon^{adfh}
                         +\varepsilon^{adeg}\varepsilon^{bcfh}+\varepsilon^{bdeg}\varepsilon^{acfh} \notag \\
                         & \qquad\qquad\qquad\qquad 
                         +\varepsilon^{acfg}\varepsilon^{bdeh}+\varepsilon^{bcfg}\varepsilon^{adeh}
                         +\varepsilon^{adfg}\varepsilon^{bceh}+\varepsilon^{bdfg}\varepsilon^{aceh}\Bigr), 
\\
&\xi_{IJ}^{abcdef}
=
\hat a_{IJKL}\Bigl(
 \epsilon^K{}^{ace}\epsilon^L{}^{bdf}
+\epsilon^K{}^{bce}\epsilon^L{}^{adf}
+\epsilon^K{}^{ade}\epsilon^L{}^{bcf}
+\epsilon^K{}^{bde}\epsilon^L{}^{acf}
\Bigr)
\notag \\ & \qquad\quad
+
\hat b_{IJ}\,g_{gh}\Bigl( \varepsilon^{aceg}\varepsilon^{bdfh}+\varepsilon^{bceg}\varepsilon^{adfh}
                         +\varepsilon^{adeg}\varepsilon^{bcfh}+\varepsilon^{bdeg}\varepsilon^{acfh} \notag \\
                         & \qquad\qquad\qquad\qquad 
                         +\varepsilon^{acfg}\varepsilon^{bdeh}+\varepsilon^{bcfg}\varepsilon^{adeh}
                         +\varepsilon^{adfg}\varepsilon^{bceh}+\varepsilon^{bdfg}\varepsilon^{aceh}
\Bigr), 
\\
&\xi^{abcdefgh}=
\hat a
\Big(
\varepsilon^{aceg}\varepsilon^{bdfh}+\varepsilon^{bceg}\varepsilon^{adfh}
                         +\varepsilon^{adeg}\varepsilon^{bcfh}+\varepsilon^{bdeg}\varepsilon^{acfh} \notag \\
                         & \qquad\qquad\qquad\qquad
+\varepsilon^{acfg}\varepsilon^{bdeh}+\varepsilon^{bcfg}\varepsilon^{adeh}
                         +\varepsilon^{adfg}\varepsilon^{bceh}+\varepsilon^{bdfg}\varepsilon^{aceh}
\Big)
,
\end{align}
}%
where $\hat a$, $\hat a_{IJ}$, $\hat a_{IJKL}$, $\hat b$ and $\hat
b_{IJ}$ are arbitrary functions of $\phi^M$ and $X^{NO}$ satisfying
$\hat a_{IJ} = \hat a_{JI}$ and $\hat a_{IJKL} = \hat a_{IJLK}$.
Here we have defined $\epsilon^I{}^{abc}$ as 
\begin{align}
\epsilon^I{}^{abc} = \varepsilon^{abcd}\phi^I_{|d}.
\end{align}

\section{Explicit form of ${\cal Q}_I$}
\label{App:Q_I}

\allowdisplaybreaks


The explicit form of ${\cal Q}_I$ is given as follows:
\begin{align}
{\cal Q}_{I}\equiv {\cal Q}_I^{(A)}+{\cal Q}_I^{(B)}+{\cal Q}_I^{(C)}+{\cal Q}_I^{(D)}+{\cal Q}_I^{(E)}+{\cal Q}_I^{(G)}+{\cal Q}_I^{(H)}+{\cal Q}_I^{(I)}+{\cal Q}_I^{(J)}+{\cal Q}_I^{(K)}+{\cal Q}_I^{(L)},
\end{align}
with
\begin{align}
\mathcal{Q}_I^{(A)}&=
A_{,I}, 
\\
\mathcal{Q}_I^{(B)}&=-2 B_{IJ,K} X^{JK}
-B_{IJ,KL}\phi^{K}_{|c}\phi^{L}{}^{|cb}\phi^{J}_{|b}
+ B_{IJ} \phi^J_{|b}{}^{|b},\\
\mathcal{Q}_I^{(C)}&= 
C_{J,I} \phi^J_{|c}{}^{|c},
\\
\mathcal{Q}_I^{(D)} &=
D_{JKL,I}\delta^{ce}_{df}\phi^J_{|c} \phi^K{}^{|d}\phi^L_{|e} {}^{|f}
+2D_{IJK,L}X^{JL}\phi^{K}_{|c}{}^{|c}
+D_{IJK,L}\phi^{J}{}^{|c}\phi^{K}_{|cd}\phi^L{}^{|d} 
\notag \\
& \quad
+D_{IJK,LM} \delta^{ce}_{bf}\phi^{L}_{|d}\phi^{M}{}^{|db}\phi^J_{|c} \phi^K_{|e}{}^{|f}  
-D_{IJK}\delta^{ce}_{bf}\phi^J_{|c}{}^{|b} \phi^K_{|e}{}^{|f}
-\frac{1}{2}D_{IJK}\delta^{ce}_{bf}\phi^J_{|c} \phi^K{}^{|l}R_{el}^{~~\,bf}, 
\\
\mathcal{Q}_I^{(E)} &=
E_{JKLMN,I}\delta^{ceg}_{dfh}\phi^J_{|c} \phi^K{}^{|d}\phi^L_{|e} \phi^M{}^{|f}\phi^N_{|g}{}^{|h}
-2E_{LJKIM,N}\delta^{ceg}_{dfh}\phi^L_{|c} \phi^J{}^{|d}\phi^K_{|e} \phi^N{}^{|f}\phi^M_{|g}{}^{|h}
\notag \\ &\quad
-2 E_{LJKIM,NO}\delta^{ceg}_{bfh}\phi^{N}_{|l}\phi^{O}{}{}^{|lb}\phi^L_{|c} \phi^K_{|e} \phi^J{}^{|f}\phi^M_{|g}{}^{|h}
+4E_{LJKIM}\delta^{ceg}_{bfh}\phi^L_{|c}{}^{|b} \phi^K_{|e} \phi^J{}^{|f}\phi^M_{|g} {}^{|h} 
\notag \\ &\quad
+ E_{LJKIM}\delta^{ceg}_{bfh}\phi^L_{|c} \phi^K_{|e} \phi^J{}^{|f}\phi^M{}^{|l}R_{gl}^{~\,bh},
\\
\mathcal{Q}_I^{(G)}&=
G_{JK,I}
\delta^{ce}_{df}\phi^J_{|c}{}^{|d} \phi^K_{|e}{}^{|f}, 
\\
\mathcal{Q}_I^{(H)} 
&=
H_{JKLM,I} \delta^{ceg}_{dfh}\phi^J_{|c}\phi^K{}^{|d} \phi^L_{|e}{}^{|f}\phi^M_{|g}{}^{|h} 
-
H_{IJKL}\delta^{ceg}_{bfh}\phi^J_{|c}{}^{|b} \phi^K_{|e}{}^{|f}\phi^L_{|g}{}^{|h} \notag \\
&\quad+2H_{IJKL,M}X^{JM}\delta^{eg}_{fh} \phi^K_{|e}{}^{|f}\phi^L_{|g}{}^{|h}
+2
H_{IJKL,M}\delta^{cg}_{fh}\phi^J_{|c} \phi^M{}^{|e}\phi^K_{|e}{}^{|f}\phi^L_{|g}{}^{|h}
\notag \\& \quad
-
H_{IJKL}\delta^{ceg}_{bfh}\phi^J_{|c} \phi^K_{|e}{}^{|f}\phi^L{}^{|l}R_{gl}^{~~\,bh}
+H_{IJKL,MN}\delta^{ceg}_{bfh}\phi^{M}_{|l}\phi^{N}{}^{|lb}\phi^J_{|c} \phi^K_{|e}{}^{|f}\phi^L_{|g}{}^{|h},
\\
\mathcal{Q}_I^{(I)}&=
I_{,I}
\delta^{ce}_{df}R_{ce}^{~~\,df}, 
\\
\mathcal{Q}_I^{(J)} 
&=
J_{JK,I}\delta^{ceg}_{dfh}\phi^J_{|c}\phi^K{}^{|d}R_{eg}^{~~\,fh}
+2J_{IJ,K}X^{JK}\delta^{eg}_{fh}R_{eg}^{~~\,fh}
+2
J_{IJ,K}
\delta^{cg}_{fh}\phi^J_{|c}\phi^{K}{}^{|e}R_{eg}^{~~\,fh} 
\notag \\&\quad
+J_{IJ,KL}\delta^{ceg}_{bfh}\phi^{K}_{|d}\phi^{L}{}^{|db}\phi^J_{|c}R_{eg}^{~~\,fh}
-
J_{IJ}\delta^{ceg}_{bfh}\phi^J_{|c}{}^{|b}R_{eg}^{~~\,fh},
\\
\mathcal{Q}_I^{(K)}&=
K_{J,I}
\delta^{ceg}_{dfh}\phi^J_{|c}{}^{|d}R_{eg}^{~~\,fh}
-\frac12 K_{J,IK}
\delta^{cegl}_{dfhm}\phi^K_{|c}{}^{|d}\phi^J_{|e}{}^{|f}R_{gl}^{~~\,hm} 
-\frac{1}{8}K_I\delta_{dfhm}^{cegl}R_{ce}^{~~\,df}R_{gl}^{~~\,hm},
\\
\mathcal{Q}_I^{(L)}&=
L_{JKL,I}\delta^{ceg}_{dfh}\phi^J_{|c}{}^{|d}\phi^K_{|e}{}^{|f}\phi^L_{|g}{}^{|h} 
-\frac14 L_{LJK,IM}
\delta^{cegl}_{dfhm}\phi^M_{|c}{}^{|d}\phi^L_{|e}{}^{|f}\phi^J_{|g}{}^{|h}\phi^K_{|l}{}^{|m}.
\end{align}

\section{Euler-Lagrange Equations}
\label{App:ELeq}

The explicit forms of the Euler-Lagrange equations from $\mathcal{L}_{1}$--$\mathcal{L}_{7}$
are given as follows:
\begin{align}
E^{ab}\left( \mathcal{L}_1 \right) 
&=
M^{(1)}_{I,J}
\left( \phi^I{}^{|(a}\phi^J{}^{|b)}
+g^{ab}X^{IJ}
\right) 
-\frac12M^{(1)}_{I,JK}
\left( g^{l(b}\delta^{a)ce}_{ldf}\phi^I_{|c}\phi^J{}^{|d}\phi^K_{|e}{}^{|f}
+2X^{IJ}
g^{l(b}\delta^{a)c}_{ld}\phi^K_{|c}{}^{|d}\right),
\\
E^{ab}\left( \mathcal{L}_2\right)
&=
 2 M^{(2)}_{,I,J} \left( \phi^I{}^{|(a}\phi^J{}^{|b)}+2 g^{ab} X^{IJ} \right) 
 -2\left( M^{(2)}_{,I} + 2 M^{(2)}_{,K,IJ} X^{JK} \right) g^{l(a}\delta^{b)c}_{ld}\phi^{I}_{|c}{}^{|d}
\notag \\
& \quad -\frac12 M^{(2)}_{,IJ}\, g^{l(a}\delta^{b)ceg}_{ldfh}\phi^I_{|c}\phi^J{}^{|d}R_{eg}^{~~\,fh}
+\left(\frac12 M^{(2)} - M^{(2)}_{,IJ} X^{IJ} \right) g^{l(a} \delta^{b)ce}_{ldf}R_{ce}^{~~\,df}
\notag \\ & \quad
-\left(  M^{(2)}_{,IJ} + 2 M^{(2)}_{,IJ,KL}X^{KL} \right)g^{l(a}\delta^{b)ce}_{ldf}\phi^{I}_{|c}{}^{|d} \phi^{J}_{|e}{}^{|f}
- M^{(2)}_{,IJ,KL} g^{l(a}\delta^{b)ceg}_{ldfh}\phi^I_{|c}\phi^J{}^{|d}\phi^K_{|e}{}^{|f}\phi^L_{|g}{}^{|h}
\notag \\ &\quad
-4 M^{(2)}_{,I,JK} \, g^{l(a}\delta^{b)ce}_{ldf}\phi^{I}_{|c}\phi^J{}^{|d}\phi^K_{|e}{}^{|f},
\\
E^{ab}\left( \mathcal{L}_3\right)
&=
-2 M^{(3)}_{IJK,L} X^{IJ} X^{KL}\,g^{ab} 
-M^{(3)}_{IJK,L} \left(
X^{KL}\phi^I{}^{|(a}\phi^J{}^{|b)}+2X^{IJ}\phi^K{}^{|(a}\phi^L{}^{|b)} 
+ 2 X^{L(I}\phi^{J)}{}^{|(a}\phi^K{}^{|b)}
\right)
\notag \\ & \quad
+\frac32 M^{(3)}_{(IJK)} 
\left(
g^{l(a} \delta^{b)ce}_{ldf}\phi^I_{|c}\phi^J{}^{|d}\phi^K_{|e}{}^{|f}
+ 2 X^{IJ}\, g^{l(a} \delta^{b)c}_{ld}\phi^K_{|c}{}^{|d}
\right)
-\frac12 M^{(3)}_{IJK,LM} \, g^{l(a}\delta^{b)ceg}_{ldfh}\phi^I_{|c}\phi^J{}^{|d}\phi^K_{|e}\phi^L{}^{|f}\phi^M_{|g}{}^{|h}
\notag \\ & \quad
+
\left( M^{(3)}_{ILM,JK} - 2 M^{(3)}_{I[JM],KL} \right)
X^{LM} \, g^{l(a}\delta^{b)cg}_{lfh}\phi^I_{|c}\phi^J{}^{|f}\phi^K_{|g}{}^{|h} 
+ 2 M^{(3)}_{IJK,LM} X^{IL} X^{JK}\, g^{l(a}\delta^{b)c}_{ld}\phi^M_{|c}{}^{|d},
\\
E^{ab}\left( \mathcal{L}_4\right) 
&=
-
M^{(4)}_{I,JK} X^{IJ}\, g^{n(a}\delta^{b)egl}_{nfhm}\phi^K_{|e}{}^{|f}R_{gl}^{~~\,hm} 
- M^{(4)}_{I,J} X^{IJ} \, g^{l(a} \delta^{b)eg}_{lfh}R_{eg}^{~~\,fh}
- M^{(4)}_{I,J} \,g^{l(a} \delta^{b)ceg}_{ldfh}\phi^I_{|c}\phi^J{}^{|d}R_{eg}^{~~\,fh}
\notag \\ & \quad
-\frac23 \left( M^{(4)}_{I,JK,LM} X^{IJ} + M^{(4)}_{K,LM} \right) g^{n(a} \delta^{b)egl}_{nfhm}
\phi^K_{|e}{}^{|f}\phi^L_{|g}{}^{|h}\phi^M_{|l}{}^{|m}
-2 M^{(4)}_{I,J,K} \, g^{l(a} \delta^{b)ce}_{ldf}\phi^J_{|c}\phi^K{}^{|d}\phi^I_{|e}{}^{|f}
\notag \\ & \quad
- 2 \left( M^{(4)}_{I,J} + M^{(4)}_{K,IJ,L} X^{KL} \right) 
g^{l(a} \delta^{b)ce}_{ldf}\phi^I_{|c}{}^{|d}\phi^J_{|e}{}^{|f} 
- 2 M^{(4)}_{I,K(J,L)} \, g^{l(a} \delta^{b)ceg}_{ldfh}\phi^J_{|c}\phi^L{}^{|d}\phi^I_{|e}{}^{|f}\phi^K_{|g}{}^{|h}
,
\\
E^{ab}\left( \mathcal{L}_5\right) 
&=
\left( M^{(5)}_{IJ} + 2 M^{(5)}_{IK,JL} X^{KL} \right)X^{IJ}\, g^{l(a} \delta^{b)eg}_{lfh}R_{eg}^{~~\,fh}
\notag \\ & \quad
+\left[ M^{(5)}_{IJ} + \left( 2 M^{(5)}_{IK,JL} - M^{(5)}_{IJ,KL} \right) X^{KL} 
\right] g^{l(a} \delta^{b)ceg}_{ldfh}\phi^I_{|c}\phi^J{}^{|d}R_{eg}^{~~\,fh}
\notag \\ & \quad
+ 2 \left[
M^{(5)}_{IJ}+  \left(
M^{(5)}_{KL,IJ} 
+ 4 M^{(5)}_{IK,JL} \right) X^{KL} + 2 M^{(5)}_{MN,KL,IJ} X^{KM} X^{LN}
\right]
g^{l(a} \delta^{b)ce}_{ldf} \phi^I_{|c}{}^{|d}\phi^J_{|e}{}^{|f}
\notag \\ & \quad
+2 \left[
2 M^{(5)}_{IK,JL} + \left(
2 M^{(5)}_{IM,JN,KL} - M^{(5)}_{IJ,KL,MN}
\right) X^{MN}
\right]
g^{l(a} \delta^{b)ceg}_{ldfh}\phi^I_{|c}\phi^J{}^{|d}\phi^K_{|e}{}^{|f}\phi^L_{|g}{}^{|h}
\notag \\ & \quad
+ 2 \left[ 
 2 M^{(5)}_{KI,J} - M^{(5)}_{IJ,K}  + 2 \left( 2 M^{(5)}_{IL,J,MK} - M^{(5)}_{IJ,L,MK} \right) X^{LM} 
\right] g^{l(a} \delta^{b)ce}_{ldf}\phi^I_{|c}\phi^J{}^{|d}\phi^K_{|e}{}^{|f}
\notag \\ & \quad
- 16 M^{(5)}_{I[J,L],K} X^{IJ} X^{KL} g^{ab}
- 4 \left(
M^{(5)}_{IJ,K,L} + M^{(5)}_{KL,I,J} - 2 M^{(5)}_{IK,J,L}
\right) X^{KL} \phi^I{}^{|(a} \phi^{J}{}^{|b)}
\notag \\ &\quad
-4 M^{(5)}_{IJ,K,LM} g^{l(a} \delta^{b)ceg}_{ldfh}\phi^I_{|c}\phi^J{}^{|d}\phi^K_{|e}\phi^L{}^{|f}\phi^M_{|g}{}^{|h},
\\
E^{ab}\left( \mathcal{L}_6\right)
&=
\frac{1}{2}M^{(6)}g^{ab}
+
\frac12  M^{(6)}_{,IJ} \phi^I{}^{|(a}\phi^J{}^{|b)},
\\
E^{ab}\left( \mathcal{L}_7 \right)
&=
\left(
M^{(7)}_{IJKLM}+M^{(7)}_{IJMLK}+M^{(7)}_{IJKML}
\right)g^{l(a}\delta^{b)ceg}_{ldfh}\phi^I_{|c}\phi^J{}^{|d}\phi^K_{|e}\phi^L{}^{|f}\phi^M_{|g}{}^{|h} 
\notag \\ & \quad
+ 4 \left(
M^{(7)}_{IJKLM}+M^{(7)}_{KLMIJ}+M^{(7)}_{KLIMJ}
\right) X^{IJ} \, g^{l(a}\delta^{b)ce}_{ldf}\phi^K_{|c}\phi^L{}^{|d}\phi^M_{|e}{}^{|f}
\notag \\ & \quad
+ \left(
-M^{(7)}_{IJKLN,OM} + 2 M^{(7)}_{IJNLK,OM} + 2 M^{(7)}_{INKLO,JM}
\right) X^{NO} \, g^{l(a} \delta^{b)ceg}_{ldfh}\phi^I_{|c}\phi^J{}^{|d}\phi^K_{|e}\phi^L{}^{|f}\phi^M_{|g}{}^{|h} 
\notag \\ & \quad
+ 8 M^{(7)}_{NLIJM,KO}  X^{LM} X^{NO} \, g^{l(a}\delta^{b)ce}_{ldf}\phi^I_{|c}\phi^J{}^{|d}\phi^K_{|e}{}^{|f}
+ 8 M^{(7)}_{IJKLM,N} X^{IJ} X^{KL} X^{MN} \,g^{ab}
\notag \\ & \quad
+ 8 \left( M^{(7)}_{IJKLM,N} - 2M^{(7)}_{IJ(NK)M,L} + M^{(7)}_{IJMNK,L} \right) X^{IJ} X^{KL} \phi^{M}{}^{|(a}\phi^N{}^{|b)}.
\end{align}

\section{Construction of Lagrangian: single scalar-field case}
\label{App:CLS}

In this appendix we briefly review the construction of the Lagrangian
for the most general {\it single} scalar-tensor theory with second
order equations of motion (see Ref.~\cite{Horndeski:1974wa} for more
detail). The most general second order field equations in the
single scalar-field case are given by
\begin{align}
\calG{}^{a}_{b}&=
A\delta^{a}_{b}
+\left( -2 \mathcal{F}^{\prime \prime}  + \dot{A}
+2 
D'X
\right)\phi^{|a} \phi_{|b}
+\left(
-2 \mathcal{F}^{\prime} +2D  X
\right) \delta^{ac}_{bd}\phi^{|d}_{|c}
\notag \\&
+ D\delta^{ace}_{bdf}\phi_{|c} \phi^{|d}\phi^{|f}_{|e}
+ \frac{1}{2}\mathcal{F} \delta^{ace}_{bdf}R_{ce}^{~~\,df}
+ \dot{\mathcal{F}} 
\delta^{ace}_{bdf}\phi^{|d}_{|c} \phi^{|f}_{|e}
\notag \\&
+
J\delta^{aceg}_{bdfh}\phi_{|c}\phi^{ |d}R_{eg}^{~~\,fh}
+ 2 \dot{J}\delta^{aceg}_{bdfh}\phi_{|c} \phi^{|d}\phi^{ |f}_{|e}\phi^{|h}_{|g}
+K\delta^{aceg}_{bdfh}\phi^{|d}_{|c}R_{eg}^{~~\,fh}
+\frac23 \dot{K}
\delta^{aceg}_{bdfh}\phi^{|d}_{|c} \phi^{|f}_{|e}\phi^{|h}_{|g} ,\label{singleEOM}
\end{align}
where $\phi$ is a scalar field and $X=-\left( \partial \phi
\right)^2/2$.  A prime ``\;$\prime$\;'' and a dot ``\;$\dot{}$\;''
denote derivatives with respect to $\phi$ and $X$, respectively, and
$A,D,J$, and $K$ are arbitrary functions of $\phi$ and $X$, while
$\mathcal{F}$ is related to the other functions as
\begin{align}
\mathcal{F}=\int \left( 2J-2K^{\prime}+4\dot{J}X\right)dX +\mathcal{W},
\end{align}
where $\mathcal{W}$ is an arbitrary function of $\phi$. 
As explained in the main text, candidates for the Lagrangian can be guessed from the trace of the field equations as 
\begin{align}
\mathcal{L}_1&=
 \sqrt{-g} \, 
M^{(1)}\phi^{|c}_{|c},\label{singLagM1} \\
\mathcal{L}_2&= 
\sqrt{-g} 
\left( M^{(2)}{}\delta^{ce}_{df}R_{ce}^{~~\,df}
+2 \dot{M}^{(2)}
\delta^{ce}_{df}\phi^{|d}_{|c} \phi^{|f}_{|e}
\right),\\
\mathcal{L}_3&=
\sqrt{-g} \, M^{(3)}\delta^{ce}_{df}\phi_{|c}\phi^{|d}\phi^{|f}_{|e},\\
\mathcal{L}_4&=
\sqrt{-g} 
\left( 
M^{(4)}\delta^{ceg}_{dfh}\phi_{|c}^{|d}R_{eg}^{~~\,fh}
+\frac{2}{3} \dot{M}^{(4)} \delta^{ceg}_{dfh}\phi^{|d}_{|c} \phi^{|f}_{|e} \phi^{|h}_{|g}
\right),\\
\mathcal{L}_5&=
\sqrt{-g} 
\left( 
M^{(5)}\delta^{ceg}_{dfh}\phi_{|c}\phi^{|d}R_{eg}^{~~\,fh}
+2 \dot{M}^{(5)}\delta^{ceg}_{dfh}\phi_{|c}\phi^{|d}\phi^{|f}_{|e}\phi^{|h}_{|g}
\right), \\
\mathcal{L}_{6}&=\sqrt{-g} \, M^{(6)}.\label{singLagM6} 
\end{align} 
By comparing the most general field equations (\ref{singleEOM}) with
the Euler-Lagrange equations obtained from (\ref{singLagM1})--(\ref{singLagM6}),
it can be seen that the free functions are related as
\begin{align}
A&=
 M^{(1)}{}^{\prime} X
+ 4 M^{(2)}{}^{\prime \prime}X
- 2 M^{(3)}{}^{\prime} X^2
+\frac{1}{2}M^{(6)} \label{singA},
\\
D&=
-\frac12 \dot{M}^{(1)}
-4 \dot{M}^{(2)}{}^{\prime}
+\frac32 M^{(3)}
+  \dot{M}^{(3)} X
-2 M^{(4)}{}^{\prime \prime}+2M^{(5)}{}^{\prime}+4\dot{M}^{(5)}{}^{\prime}X \label{singD},
\\
\mathcal{F}&=
M^{(2)}
-2 \dot{M}^{(2)}X
-2 M^{(4)}{}^{\prime}X
+2{M}^{(5)}X
+4 \dot{M}^{(5)}X^2,\label{singF}
\\
J&=
-\frac12 \dot{M}^{(2)}
- M^{(4)}{}^{\prime}
+M^{(5)}
+ \dot{M}^{(5)}X, \label{singJ}
\\
K&=
-\dot{M}^{(4)}X, \label{singK}\\
-2 \mathcal{F}^{\prime \prime}
+ \dot{A}
+ 2 D^{\prime}X
&= 
M^{(1)}{}^{\prime}
+ 2 M^{(2)}{}^{\prime \prime}
-M^{(3)}{}^{\prime} X+\frac12 \dot{M}^{(6)}, \label{singddF} \\
-2 \mathcal{F}^{\prime}
+2  D X
&
=
 -\dot{M}^{(1)}X
-2 \left( M^{(2)}{}^{\prime} + 2 \dot{M}^{(2)}X \right)
+ 3 M^{(3)}X
+2 \dot{M}^{(3)} X^2 \label{singdF}.
\end{align}

In order to identify the Lagrangian,
we need to solve the above equations for $M^{(1)}$--$M^{(6)}$.
First, by integrating Eq.~(\ref{singK}) we obtain
\begin{equation}
M^{(4)}=-\int \frac{K}{X}dX.\label{singM4}
\end{equation}
Substituting Eq.~(\ref{singM4}) into Eqs.~(\ref{singF}) and~(\ref{singJ}), we then find
\begin{align}
M^{(2)}&=-2\int \left( M^{(4)}{}^{\prime} -M^{(5)}-2X\dot{M}^{(5)}\right) dX ,\label{singM2}\\
M^{(5)}&=-\int \frac{J}{X}dX
\label{singM5}.
\end{align}
Finally, we can integrate
Eqs.~(\ref{singA}),~(\ref{singD}),~(\ref{singddF}), and~(\ref{singdF}) to
identify $M^{(1)}$, $M^{(3)}$, and $M^{(6)}$ with the help of
Eqs.~(\ref{singM4}),~(\ref{singM2}) and~(\ref{singM5}), leading to
\begin{align}
M^{(1)}&=-3\left( 2M^{(2)}{}^{\prime}-XM^{(3)}\right), \label{singM1}\\
M^{(3)}&=-2\int \frac{D}{X}dX, 
\label{singM3}
\end{align}
and
\begin{align}
M^{(6)}=
2A
+4XM^{(2)}{}^{\prime \prime}-2M^{(3)}{}^{\prime}. \label{singM6}
\end{align}


\end{document}